\documentclass[aip,amsmath,amssymb,reprint,nofootinbib,floatfix]{revtex4-1}
\usepackage{graphicx}
\usepackage{dcolumn}
\usepackage{bm}
\usepackage[utf8]{inputenc}
\usepackage[T1]{fontenc}
\usepackage{mathptmx}
\usepackage{comment}

\preprint{AIP/123-QED}
\usepackage{lmodern,graphicx}
\usepackage{times}
\usepackage{mathptmx}
\usepackage{epstopdf}
\usepackage{xcolor}
\usepackage[colorlinks,linkcolor=red,urlcolor=blue,citecolor=red]{hyperref}
\usepackage[normalem]{ulem}
\usepackage{xspace}
\usepackage{float}
\hypersetup{
    colorlinks=true,
    linkcolor=blue,
    filecolor=magenta,      
    urlcolor=cyan,
    citecolor=cyan
}
\usepackage{array}
\newcolumntype{P}[1]{>{\centering\arraybackslash}p{#1}}
\begin{document}

\title[]{Synchronization transition in the two-dimensional Kuramoto model with dichotomous noise}

\author{Mrinal Sarkar}
\affiliation{ 
Department of Physics, Indian Institute of Technology Madras, Chennai-600036, India.
}%
\email{mrinal@physics.iitm.ac.in}


\begin{abstract}
We numerically study the celebrated Kuramoto model of identical oscillators arranged on the sites of a two-dimensional periodic square lattice and subject to nearest neighbor interactions and dichotomous noise. In the nonequilibrium stationary state attained at long time, the model exhibits a Berezinskii-Kosterlitz-Thouless ($BKT$)-like transition between a phase at low noise amplitude characterized by quasi long-range order (critically ordered phase) and algebraic decay of correlations and a phase at high noise amplitude that is characterized by complete disorder and exponential decay of correlations. The interplay between the noise amplitude and the noise correlation time is investigated, and the complete, nonequilibrium stationary-state phase diagram of the model is obtained. We further study the dynamics of a single topological defect for various amplitude and correlation time of the noise. Our analysis reveals that a finite correlation time promotes vortex excitations, thereby lowering the critical noise amplitude of the transition with an increase in correlation time. In the suitable limit, the resulting phase diagram allows to estimate the critical temperature of the equilibrium $BKT$ transition, which is consistent with that obtained from the study of the dynamics in the Gaussian white noise limit.
\end{abstract}
\maketitle

\begin{quotation}
Study of synchronization in complex systems consisting of a large population of interacting degrees of freedom and evolving in presence of stochastic force has been a subject of wide interest. Here, we explore such phenomenon within the ambit of the 2D Kuramoto model of identical phase oscillators driven by dichotomous Markov noise, which leads the system to settle into a nonequilibrium stationary state in long time. We show that in such a state, the model exhibits a phase transition analogous to the $BKT$ transition in the equilibrium stationary state of the 2D $XY$ model, whereby the system makes a transition between a phase with quasi long-range order observed at low noise amplitude, and a phase with complete disorder at high noise amplitude. A detailed investigation of the interplay between the noise amplitude and the noise correlation time has been conducted. As the topological defects play an important role in causing the phase transition, dynamics of a single topological defect is also studied. Our analysis reveals that a finite correlation time induces vortex excitations in the system and consequently, with an increase in noise correlation time the critical noise amplitude of the transition decreases. As a special case, we recover the critical temperature of the equilibrium $BKT$ transition from a study of a suitable limiting case of the dynamics. This equilibrium critical point is also estimated by extrapolating the line of transition in the phase diagram, which is consistent with that obtained from former study.

\end{quotation}


\section{Introduction}
\label{sec:intro}

Stochasticity is an inevitable characteristic of most dynamical systems observed in nature. It may be intrinsic to some systems, e.g., in biological systems that may be modeled in terms of interacting oscillators, where the natural frequencies of the oscillators may have a fluctuating part. In thermodynamic systems, it arises from  thermal fluctuations present in the system due to its interaction with the environment. To incorporate the effect of stochasticity into the dynamics, one usual way is to model it as a $\delta$-correlated Gaussian ``white noise''. This approximation holds only when the time scale of fluctuations is much shorter than the one characterizing the deterministic part of the dynamics in question. However, in many situations of interest, these two time scales are comparable to each other, and consequently, the Gaussian white noise modeling turns out to be unreasonable.

Here we aim to study the impact of a colored noise, namely, zero mean and exponentially correlated symmetric two-state Dichotomous Markov process in the framework of the so-called Kuramoto model. Over the years, this model has served as a paradigm to study analytically the phenomenon of collective synchronization \cite{kuramoto1975international,strogatz2000kuramoto}. Synchronization is one of the most fascinating emergent phenomena in complex systems consisting of a large population of interacting degrees of freedom
\cite{pikovsky2003synchronization,strogatz2004sync,gupta2018statistical}. This phenomenon is ubiquitous in different disciplines of science, including physics \cite{wiesenfeld1996synchronization, silber1993stability}, chemistry \cite{Taylor2009}, biology\cite{bier2000yeast,winfree2001geometry}, social science \cite{Nda2000} and so on. The list is quite extensive. For more examples on synchronization, we refer the article \cite{acebron2005kuramoto}.

Dichotomous noise provides a good representation of many physical and biological situations.
For instance, it models the molecular noise in genetic network arising from a single copy of a gene stochastically switching between two states (ON and OFF)\cite{potoyan2015dichotomous}. This noise can also be viewed as an external field with bounded amplitude and can be studied to compare the dynamics with other field of this category, e.g., periodically oscillating field. In appropriate limit dichotomous noise reduces to Gaussian white noise. So naturally the question arises: Does the dynamics in presence of dichotomous noise and Gaussian white noise result in same behavior?
To answer this question, let us begin with an example of the Kuramoto model. The mean-field Kuramoto model with identical natural frequencies in presence of Gaussian white noise shows a continuous phase transition between an ordered phase at low temperature and a disordered phase at high temperature at critical temperature $T_{c}=0.5$ \cite{sakaguchi1988cooperative}.  Thus in changing the temperature by a small amount near and below the transition point, the steady state order parameter also changes by a small amount. On the other hand, the same system when subject to dichotomous noise, exhibits a discontinuous transition between an ordered phase at low noise amplitude and a disordered phase at high noise amplitude \cite{tonjes2010synchronization}. Thus a small change in the noise amplitude across the transition point causes an abrupt and big jump in the steady state order parameter value, from a non-zero value to zero. The system displays hysteretic behavior while tuning the noise amplitude adiabatically in a cycle \cite{tonjes2010synchronization}. Thus the dichotomous noise changes completely the nature of the phase transition with respect to the one observed with Gaussian white noise. This highlights the fact that the dynamics of the system changes dramatically by the presence of dichotomous noise. Another example includes similar system, namely, the mean-field Kuramoto model of identical oscillators with additional onsite potential. In such systems, dichotomous noise induces creation of new phases, e.g., oscillating stationary states, hysteresis which are not observed when the system is subject to Gaussian white noise \cite{kostur2002nonequilibrium}. In general, presence of dichotomous noise is known to lead to a range of fascinating phenomena such as robust phase synchronization~\cite{roussel2001phase}, enhanced stochastic resonance~\cite{rozenfeld2000stochastic}, pattern-formation~\cite{das2013dichotomous}.
 
In fact, the dynamics in presence of dichotomous noise is fundamentally different from that of Gaussian white noise. The latter appearing in the dynamics represents thermal fluctuations arising from interaction of the system with the heat bath. The stationary state attained at long time is thermodynamic equilibrium state. The strength of the fluctuation is related to the parameter of the system via fluctuation dissipation relation. But the dichotomous noise, which is a non-Gaussian colored noise, describes nonthermal fluctuations in the system, and thus its strength is independent of the parameters of the system under study. The noise drives the system to a nonequilibrium stationary state at long time. Thus, it is not guaranteed that the phenomena observed in a system involving dichotomous noise would be same as that in presence of Gaussian white noise. In fact, it may lead to novel behaviors that are not accessible in the equilibrium system. Clearly, dichotomous noise has its own virtues and thus system driven by such noise requires an independent study, an issue we take up in the present work.

In this backdrop, we explore the impact of dichotomous noise on synchronization dynamics of a variant of the Kuramoto model, whereby the oscillators with identical frequencies and placed on the sites of a 2D periodic square lattice are interacting only with their nearest neighbors (local coupling). The resulting dynamics attains for all finite correlation time a nonequilibrium stationary state at long times. In such a state, the system displays a quasi-ordered phase at low noise amplitude and a disordered phase at high noise amplitude, thereby exhibiting a $BKT$-like transition between the two phases as one tunes the noise amplitude. We demonstrate here for the first time a nonequilibrium $BKT$-like transition driven by non-Gaussian colored noise in the framework of the Kuramoto model. Additionally, we investigate the interplay between the noise amplitude and the correlation time of the dichotomous noise in dictating the nature of the phase transition. We further study the dynamics of a single topological defect under the influence of this noise. A single defect is found to exhibit anomalous diffusion in the sense that its mean-squared displacement (MSD) shows a linear behavior not with time $t$, instead with $t/{\ln t}$ showing a logarithmic correction to the normal diffusion. Our analysis reveals that a finite-correlation time induces vortex excitations in the nonequilibrium stationary state of the dynamics. Finally, we study the dynamics in the white noise limit and recover the critical temperature of the equilibrium $BKT$ transition. The equilibrium critical point estimated by extrapolating the line of transition in the phase diagram is in well agreement with that value, which shows the consistency of our work.

Let us note that the equilibrium dynamics of our model i.e. when subject to Gaussian white noise is known to exhibit the equilibrium $BKT$ transition \cite{kosterlitz1973ordering,kosterlitz1974critical}. Even in presence of Gaussian colored noise, namely, the Ornstein-Uhlenbeck (OU) noise, the dynamics is also expected to show similar transition in the nonequilibrium stationary state \cite{paoluzzi2018effective}. In this sense, introduction of dichotomous noise does not lead to novel behavior in our system. Despite that we report on the following novel features associated with this dichotomous noise driven nonequilibrium $BKT$-like transition, which are absent in Gaussian white or colored noise driven system. 
Firstly, in system driven by Gaussian colored noise, the $BKT$ transition temperature remains same as the equilibrium one \cite{paoluzzi2018effective}, whereas our analysis reveals that, the $BKT$-transition point indeed shifts in presence of dichotomous noise.
Secondly, in contrast to Gaussian colored noise, the dynamics in presence of dichotomous noise is found to yield the maximum value of the power-law exponent of spatial correlation exceeding the equilibrium upper bound i.e. $1/4$ \cite{paoluzzi2018effective}. This implies that when subject to dichotomous noise, the quasi-ordered phase can sustain higher level of collective excitations leading to faster decay of the spatial correlation compared to Gaussian white or colored noise.

The paper is organized as follows. In Sec.~\ref{sec:model}, we define our model of study along with a list of queries to be addressed in this work. In Sec.~\ref{sec:DN_results}, we compute various statistical quantities to characterize the transition and obtain the nonequilibrium stationary state phase diagram of the dynamics in the relevant parameter space. A qualitative analysis based on the dynamics of Topological defects is also presented. The paper ends in Sec.~\ref{sec:conclusions} with conclusions and future directions of our work. In Appendix~\ref{App_DN_noise_generation} we discuss how we have generated dichotomous noise. We provide a discussion on how the stationarity is checked in Appendix~\ref{App_stationarity}. Finally, Appendix~\ref{App_Binder_cumulant_scaling} provides the scaling theory of the Binder cumulant in continuous transitions and at $BKT$ transition in equilibrium systems.

\section{Model and Our queries}
\label{sec:model}

We consider a system of Kuramoto oscillators of identical frequencies that are arranged on the sites of a two-dimensional periodic square lattice of a total of $N \equiv L \times L$ sites, in which the oscillators interact only with their nearest neighbors. The evolution equation of the phase $\theta_{i} \in [0,2\pi)$ of the $i$-{th} oscillator \textit{in presence of noise} is then given by 
\begin{equation}
\frac{{\rm d} \theta_{i}}{\rm dt} = \omega + K  \sum_{j \in nn_{i}}  \sin(\theta_{j}- \theta_{i}) + \zeta_{i}(t),
\label{eq:eom1}
\end{equation}
where $\omega$ is the natural frequency of the $i$-th oscillator and $nn$ implies that the sum is over nearest neighbors only. Here, $K > 0$ is the strength of nearest-neighbor interaction, while $\zeta_{i}(t)$ is the \textit{noise} term. By a choice of suitable reference frame (co-rotating frame) the natural frequencies $\omega$ can be set to zero without loss of generality, and one has the resulting dynamics
\begin{equation}
\frac{{\rm d} \theta_{i}}{\rm dt} = K  \sum_{j \in nn_{i}}  \sin(\theta_{j}- \theta_{i}) + \zeta_{i}(t).
\label{eq:eom_DN1}
\end{equation}

We consider the driving force $\zeta_{i}(t) \in \{-H,+H\}$ to be a dichotomous random Markov process with equal transition rate $\lambda$ between the two states $\pm H$. The noise satisfies the properties
 \begin{equation}
 \langle \zeta_i(t) \rangle =0~ \text{and}~\\
 \langle \zeta_i(t) \zeta_{j}(t') \rangle = H^2 \delta_{ij}\exp\left(-\frac{|t-t'|}{\tau}\right),
 \label{eq:DN}
\end{equation}
where $H^2>0$ and $\tau= \frac{1}{2 \lambda} >0$ are the noise strength and the noise correlation time, respectively. On further implementing the transformations $t \to
Kt,~H \to H/K,~\tau \to K\tau, ~\text{and}~\zeta_i \to \zeta_{i}/K $, the dynamics~(\ref{eq:eom_DN1}) reduces to the following dimensionless form
\begin{equation}
\frac{{\rm d} \theta_{i}}{\rm dt} =  \sum_{j \in nn_{i}}  \sin(\theta_{j}- \theta_{i}) + \zeta_{i}(t).
\label{eq:eom_DN}
\end{equation}

Note that the dimensionless noise $\zeta_{i} (t)$ in the above equation would continue to satisfy Eq.~(\ref{eq:DN}), with the difference that the quantities $H, \tau$ are now to be considered to be dimensionless. The dichotomous noise being non-Gaussian results in the dynamics (\ref{eq:eom_DN}) violating the principle of detailed balance in the stationary state. Consequently, the latter is not a Gibbs-Boltzmann equilibrium but rather a nonequilibrium stationary state.

We note that the dynamics~(\ref{eq:eom_DN}) together with~(\ref{eq:DN}), when considered in the simultaneous limit $\tau \to 0$ and $H \to \infty$ while keeping  $H^{2}\tau \to \text{fixed and finite}$, corresponds to the dynamics of the Kuramoto model of identical oscillators in presence of Gaussian white noise, the long time dynamics of which is governed by equilibrium statistical mechanics. This system, being equivalent to the 2D $XY$ model in contact with a heat bath at temperature $T \equiv H^2 \tau$, exhibits the $BKT$ transition from a low-temperature quasi-ordered phase to a high-temperature disordered phase at the critical temperature $T_{\text{BKT}}=0.9$ \cite{kosterlitz1973ordering,kosterlitz1974critical}.
On the other hand, the corresponding equilibrium system of~(\ref{eq:eom_DN}) at zero temperature does not show a phase transition in the presence of quenched disorder\cite{strogatz1988phase, hong2005collective, lee2010vortices}.

In the light of the foregoing, we ask the following questions: Does the dynamics~(\ref{eq:eom_DN}) that involves a dichotomous noise show a transition or a crossover behavior, and what is the nature of the different possible phases? In case a phase transition is possible, what is the order of the transition? How does a finite value of a correlation time $\tau$ affect the synchronization transition? What is the interplay between the noise amplitude $H$ and the noise correlation time $\tau$ in dictating the nature of the transition? These questions, pertinent as they are, have to the best of our knowledge not been addressed before and will be the focus of current work.

%
%
\section{Results and Discussion}
\label{sec:DN_results}
\begin{figure}[]
	\centering
	\includegraphics[scale=0.2]{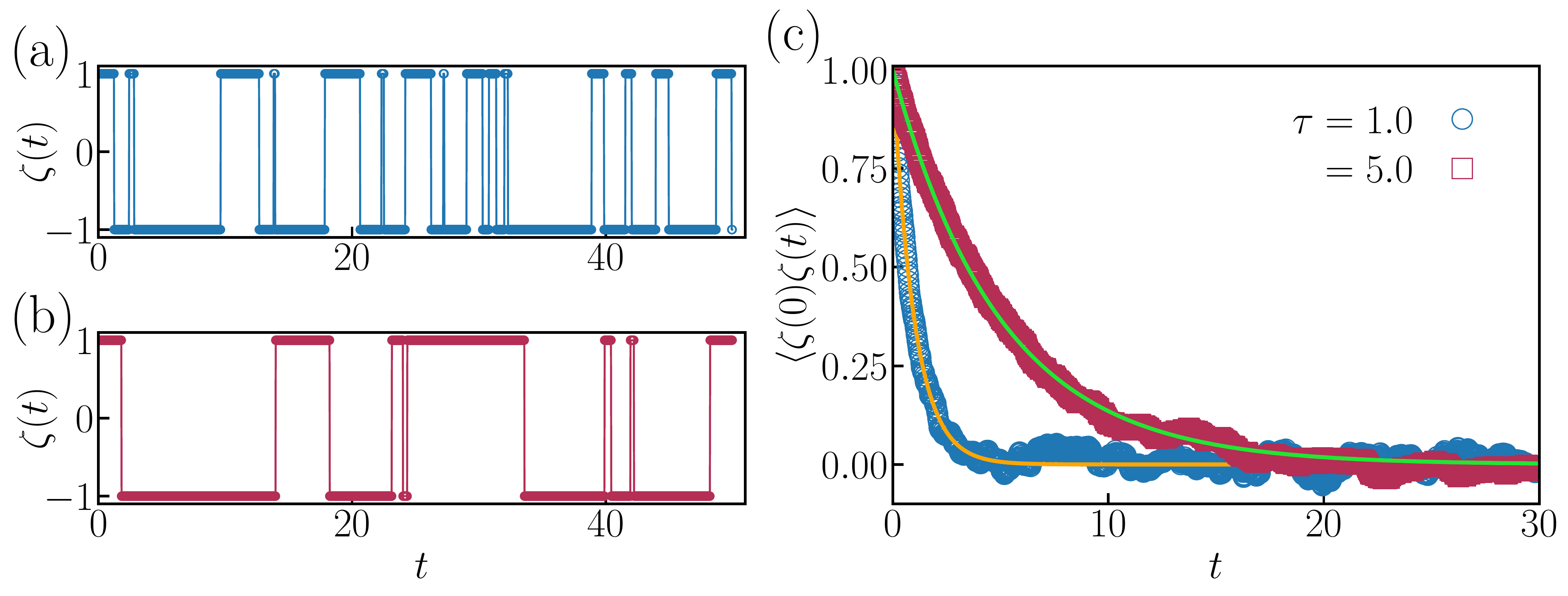}
	\caption{Typical realizations of symmetric dichotomous noise with equal transition rate between the two states $\pm H$ with $H=1$ for $\tau=1.0$ and $\tau=5.0$ are shown in (a) and (b), respectively. Fig (c) shows the corresponding auto-correlation functions denoted by blue circles and magenta squares for $\tau=1$ and $5$, respectively. The orange and green continuous lines, for $\tau=1$ and $5$ respectively, are drawn following their analytical forms as prescribed in Eq.~\ref{eq:DN}.}
	\label{fig:DN_realization}
\end{figure}

In this section, we study the behavior of various statistical quantities for our system (\ref{eq:eom_DN}). To begin with, we numerically integrate the dynamics (\ref{eq:eom_DN}) by employing fourth order Runge-Kutta method with integration time-step $dt=0.01$, with the noise satisfying the properties given by Eq.~(\ref{eq:DN}). We apply periodic boundary condition to our problem. The method to generate the dichotomous noise is provided in Appendix~\ref{App_DN_noise_generation}. We note that instead of sampling the noise from a stationary distribution, we evolve the dynamics of both the oscillator and the noise simultaneously; see Appendix~\ref{App_DN_noise_generation}. Two typical realizations of the noise for two different correlation times, namely, $\tau=1.0$ and $5.0$, along with the corresponding auto-correlation functions are shown in Fig~\ref{fig:DN_realization}. The continuous lines in Fig~\ref{fig:DN_realization}(c) correspond to the analytical form of the auto-correlation given by Eq.~(\ref{eq:DN}). In our study, the initial values of all oscillator phases are set to zero. We measure all the statistical quantities that we report in the following only after ensuring that both the Kuramoto system as well as the noise attain stationary state; please see Appendix~\ref{App_stationarity}.

\subsection{Order parameter}
The degree of phase synchronization for a system of $N$ oscillators may be quantified in terms of the usual Kuramoto synchronization order parameter~\cite{kuramoto-book}
\begin{equation}
 R e^{{\rm i} \psi} \equiv \frac{1}{N}  \sum_{j=1}^{N} e^{{\rm i} \theta_{j}},
\label{eq:order_para_definition}   
\end{equation}
where the quantity $R~(0\leq R \leq 1)$ measures the amount of synchrony present in the system at a given instant of time and $\psi$ is the average phase at that instant.

\begin{figure}[]
	\centering
	\includegraphics[scale=0.36]{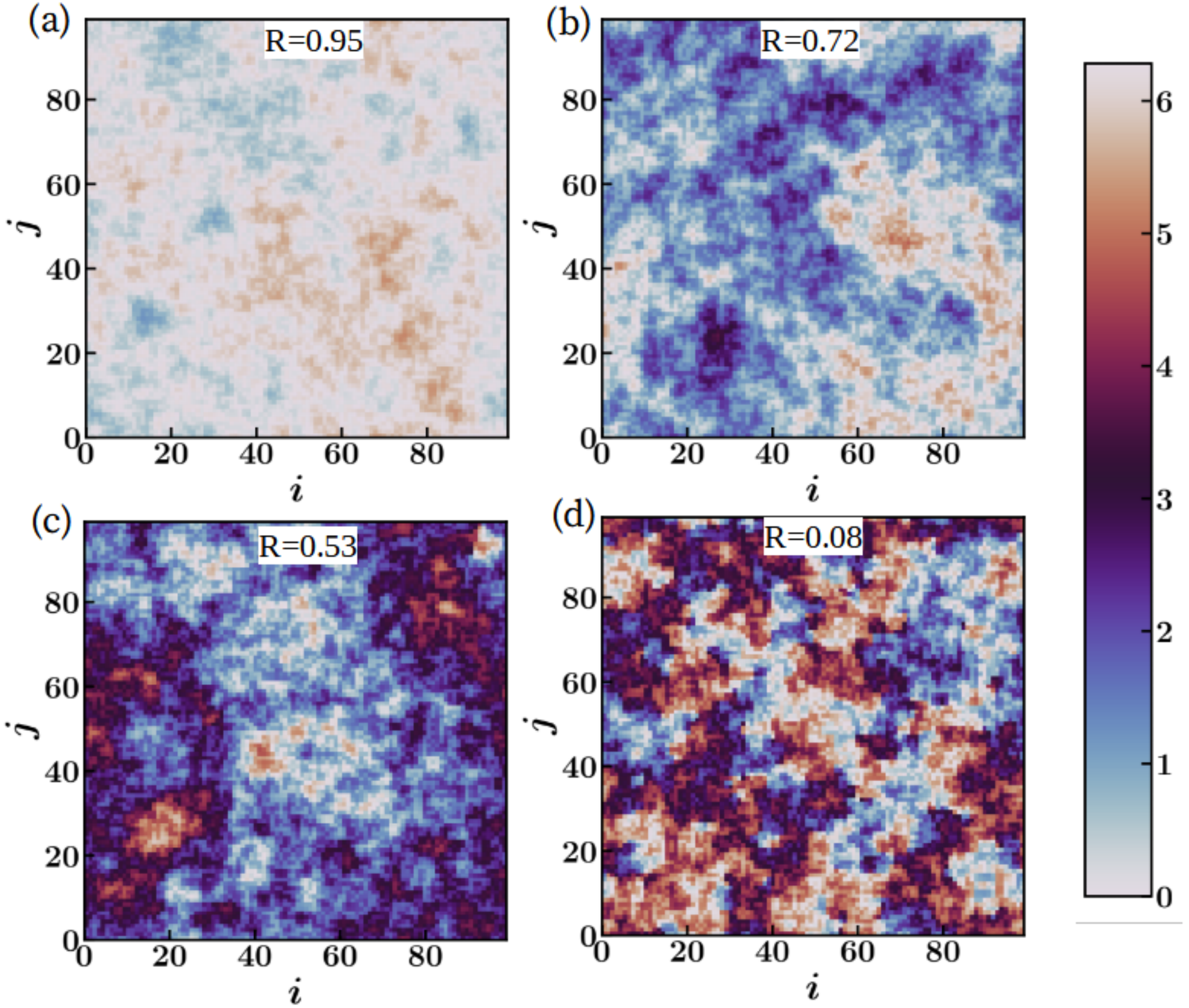}
	\caption{Shown are snapshots in the nonequilibrium stationary state of the dynamics~(\ref{eq:eom_DN}) on a lattice of size $N=100 \times 100$, at four values of dichotomous noise amplitude, namely, $H=0.5$(a), $1.0$(b), $1.28$(c)~\text{and}~$1.5$(d). Oscillators are represented by the indices ($i,j$). Each pixel represents one oscillator and the color indicates
	its phase as denoted by the color bar on the right. The noise-correlation time is chosen to be $\tau =1.0$. The corresponding values of the order parameter $R$ are also displayed. Panel (d) displays unbound vortices and anti-vortices (``Topological defects'').}
	\label{fig:DN_snapshots_all}
\end{figure}

Figure \ref{fig:DN_snapshots_all} shows time snapshots and corresponding order parameter values $R$ in the stationary state on a lattice of size $N=100\times 100$, and at four values of the dichotomous noise amplitude, namely, $H=0.5$, $1.0$, $1.28$~\text{and}~$1.5$ in panels (a), (b), (c) and (d), respectively. Each pixel represents one oscillator and the color indicates its phase as denoted by the color bar. The noise correlation time is chosen to be $\tau =1.0$. As observed from panels (a), (b) and (c), the oscillators are locally synchronized, thereby forming a cluster, whereas panel (d) displays ``Topological defects'' \textit{i.e.} unbound vortices and anti-vortices. These are the small regions in the phase-field with large phase gradient. The phase winds by integral multiple of $2 \pi$ around such defects.
We observe that for a finite $\tau$, at low value of $H$ a system of finite size shows a non-zero order parameter value in the stationary state. But this order parameter tends to zero as $N \to \infty$. This suggests that at any finite noise amplitude there is no macroscopic ordering (or equivalently, synchronized phase) in the system in the thermodynamic limit. So far as finite systems are considered, the dynamics~(\ref{eq:eom_DN}) exhibits a crossover from a low-$H$ non-zero $R$-valued phase to a high-$H$ disordered phase characterized by $R \sim \mathcal{O}(1/\sqrt{N})$. 
%

\begin{figure}[]
	\centering
	\includegraphics[scale=0.35]{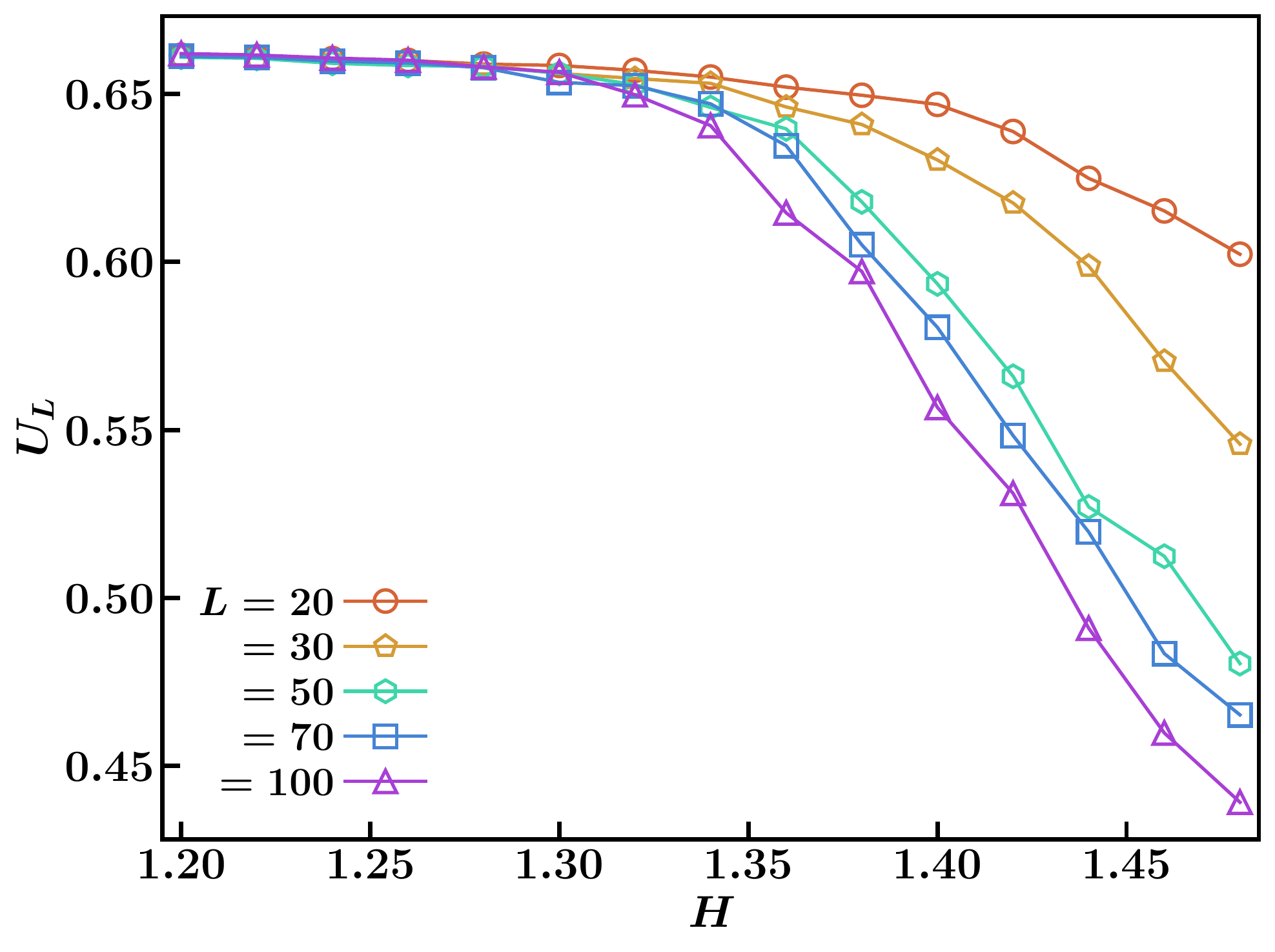}
	\caption{Variation of Binder cumulant $U_{L}$ with dichotomous noise amplitude $H$, in the nonequilibrium stationary state of the dynamics~(\ref{eq:eom_DN}), for various values of system-size $N=L\times L$ is shown. The symbols circle, pentagon, hexagon, square and triangle correspond to $U_{L}$ values for $L=20,~30,~50,~70$ and $100$, respectively. The noise-correlation time is chosen to be $\tau=1.0$. The curves for various $L$ stay collapsed up to $H_c\approx1.28$, beyond which they separate out. This indicates existence of a phase with diverging correlation length in the thermodynamic limit (critically ordered phase) in the region $H \leq H_c$.}
	\label{fig:DN_Binder_cum_plot}
\end{figure}

\subsection{Binder Cumulant}
To understand the nature of a transition as well as to locate the transition point, a useful diagnostic tool is the so-called fourth order Binder cumulant, which for a system defined on a finite lattice of linear size $L$ is given by \cite{binder1981finite, binder-book}
\begin{eqnarray}
U_{L} = 1- \left[\frac{\langle R^4 \rangle_{L}}{3\langle R^2 \rangle^{2}_{L}}\right],
\end{eqnarray}
where $\langle \cdot \rangle$ and $[\cdot]$ represent the time average in the stationary state and sample averages, respectively. Here sample average means taking average over different noise realizations.

Based on the discussion on finite-size scaling theory briefly summarized in Appendix~\ref{App_Binder_cumulant_scaling} and assuming that this scaling holds for continuous transition in a nonequilibrium system too, for large lattice sizes in the limit $L \to \infty$, one has in the ordered phase the asymptotic behavior,
$U_{L} \to 2/3$, and in the disordered phase the asymptotic behavior, $U_{L} \to 1/3$. For large but finite $L$, one has in both the phases, the correlation length $\xi$ satisfying $\xi \ll L$, and consequently, $U_{L}$ for various lattice sizes remains close to these aforementioned asymptotic values. Now, for $\xi \gg L$, the system is expected to stay close to another fixed point value $U^{*}$, independent of $L$. So, the critical parameter value at which $\xi \to \infty$ can be identified by looking for the common intersection point of the curves for $U_{L}$ vs. the relevant parameter ($H$) for lattice of various sizes.

Figure \ref{fig:DN_Binder_cum_plot} shows the variation of $U_{L}$ with noise amplitude $H$ for various values of $L$ and for a fixed value of $\tau=1.0$. The curves for various $L$ seem not to intersect at a common point but rather collapse and remain so upto a certain value ($H_{c} \approx 1.28$). On the basis of discussion in Appendix~\ref{App_Binder_cumulant_scaling}, this implies the existence of a critically ordered phase and a diverging correlation length in the range $H \leq H_{c}$. Beyond this region ($H > H_{c}$), the curves separate, suggesting the onset of disorder at higher values of $H$. We thus see that a study of the $U_{L}$ yields an estimation of $H_{c}$ as well as of the nature of the ordered phase.

\subsection{Two-point Correlation}
As discussed in the previous section, in the region $H \leq H_{c}$, the system remains in critically ordered phase i.e. $\xi$ is infinite in the thermodynamic limit, which in turn implies the power-law behavior of correlations in this phase. As a final verification, we calculate two-point first order correlation function, defined as
\begin{equation}
g^{(1)}(\vec{r},t)= \langle \cos [ \theta(\vec{r},t) - \theta(\vec{0},0) ]\rangle,
\end{equation}
where $\theta(\vec{r},t)$ represents the phase of the oscillator at position $\vec{r}$ on the lattice at time $t$, and $\langle \cdot \rangle$ represents averaging over oscillators. To study the static and dynamical properties, we investigate behavior of spatial $(g^{(1)}(r,0))$ and temporal correlation $(g^{(1)}(0,t))$ functions, respectively. The spatial correlation function $g^{(1)}(r,0)$ is computed in the following way: first circular bins are formed around an oscillator in a particular steady state configuration, the quantity $g^{(1)}(r,0)$ is calculated for each bin, and then the same process is repeated for each oscillator in that configuration, and finally, the averaging is done over all the oscillators. This whole process is repeated over sufficient number of configurations for each realization, and the quantity thus obtained is finally averaged over 100 such independent realizations. We compute the dynamical correlation function in the following way. For each realization, we let the system evolve for sufficiently long time until it reaches the stationary state. In such a state, we start our observation at a particular time instant by recording the value of the oscillators' phases, and call it $\{\theta(\vec{r},0)\}$. For each $t$ post that time instant, we calculate the quantity $g^{(1)}(0,t)$ for each oscillator, and then average over all the oscillators of the system. Finally, the quantity thus obtained is further averaged over 100 such independent realizations.

\begin{figure}[]
	\centering
	\includegraphics[scale=0.35]{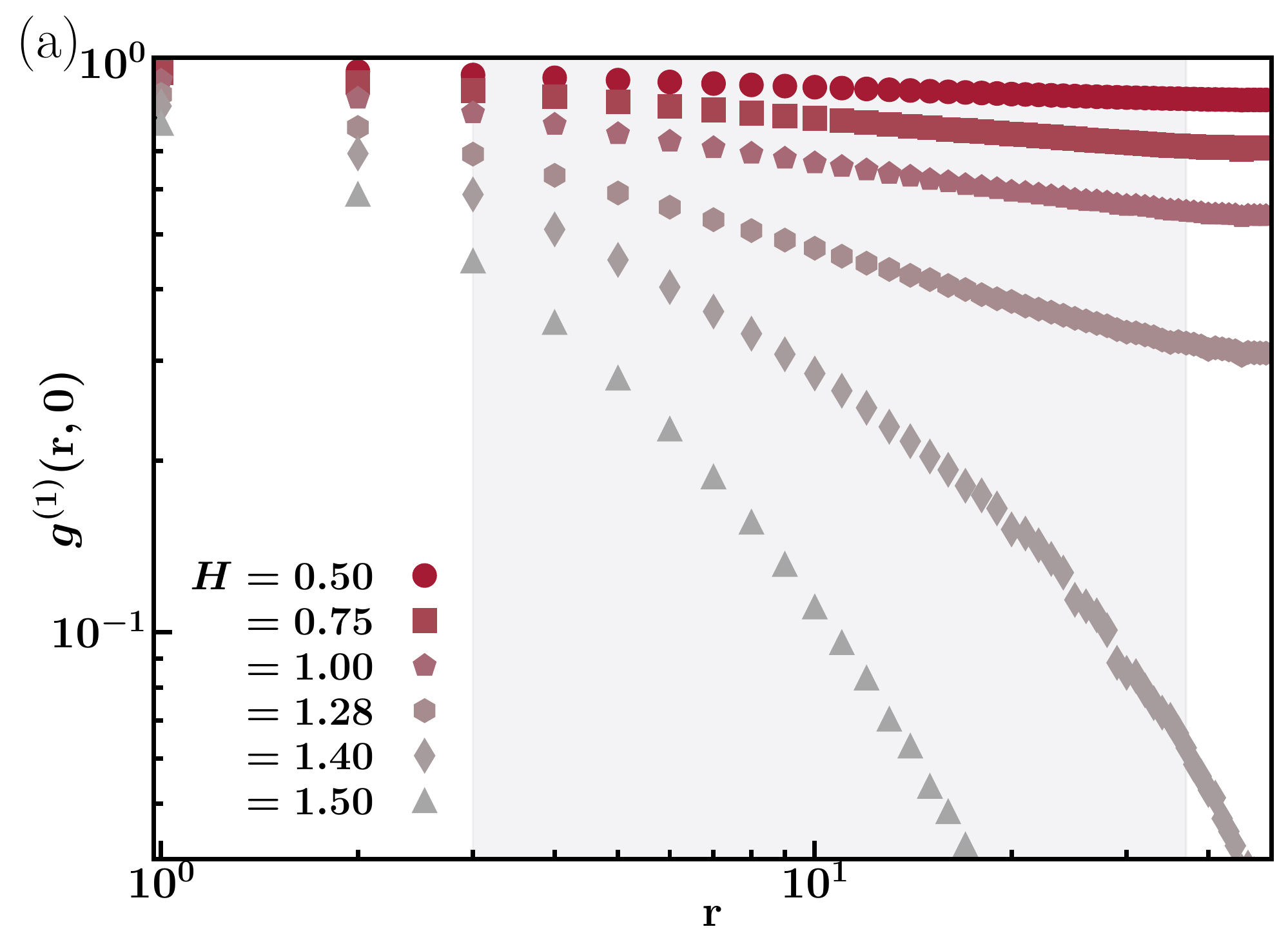}
	\includegraphics[scale=0.365]{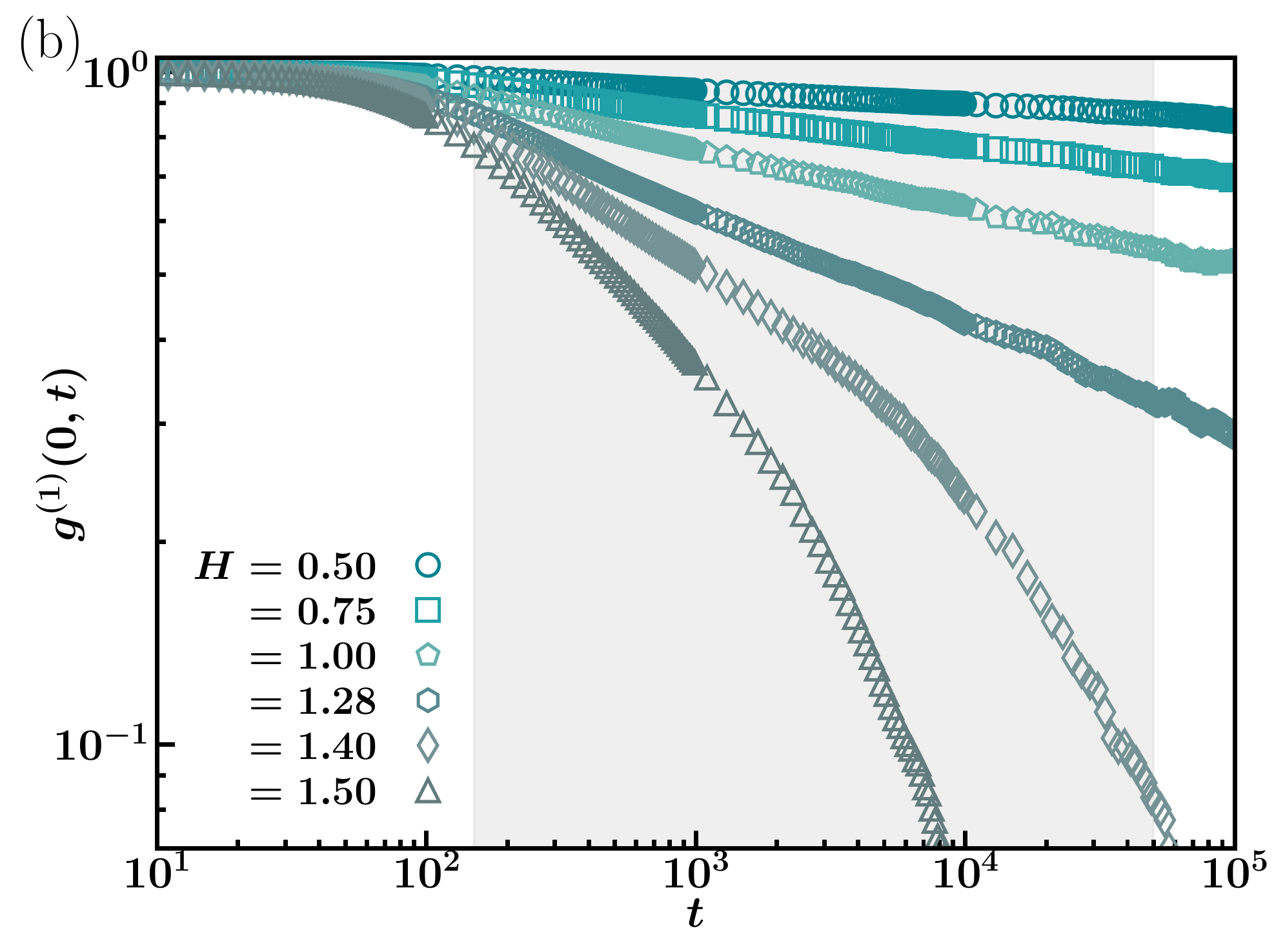}
	\caption{Spatial ($g^{(1)}(r,0)$) and dynamical ($g^{(1)}(0,t)$) correlation, on a log scale, in the nonequilibrium stationary state of the dynamics~(\ref{eq:eom_DN}) on a lattice of size $N=100\times 100$, are shown for various values of dichotomous noise-amplitudes $H$, in (a) and (b), respectively. The filled (empty) circle, square, pentagon, hexagon, diamond and triangle correspond to spatial (dynamical) correlation for $H=0.50,~0.75,~1.00,~1.28,~1.40$ and $1.50$, respectively. The noise-correlation time is chosen to be $\tau = 1.0$. The distance $r$ is in units of lattice spacing and time $t$ is in integration time steps. Both the correlation functions behave algebraically upto critical $H$-value, $H_{c} = 1.28$ and beyond that the decay is exponential. The correlation behavior was characterized over the shaded region only. }
	\label{fig:DN_corr}
\end{figure}

\begin{figure}[]
	\centering
	\includegraphics[scale=0.33]{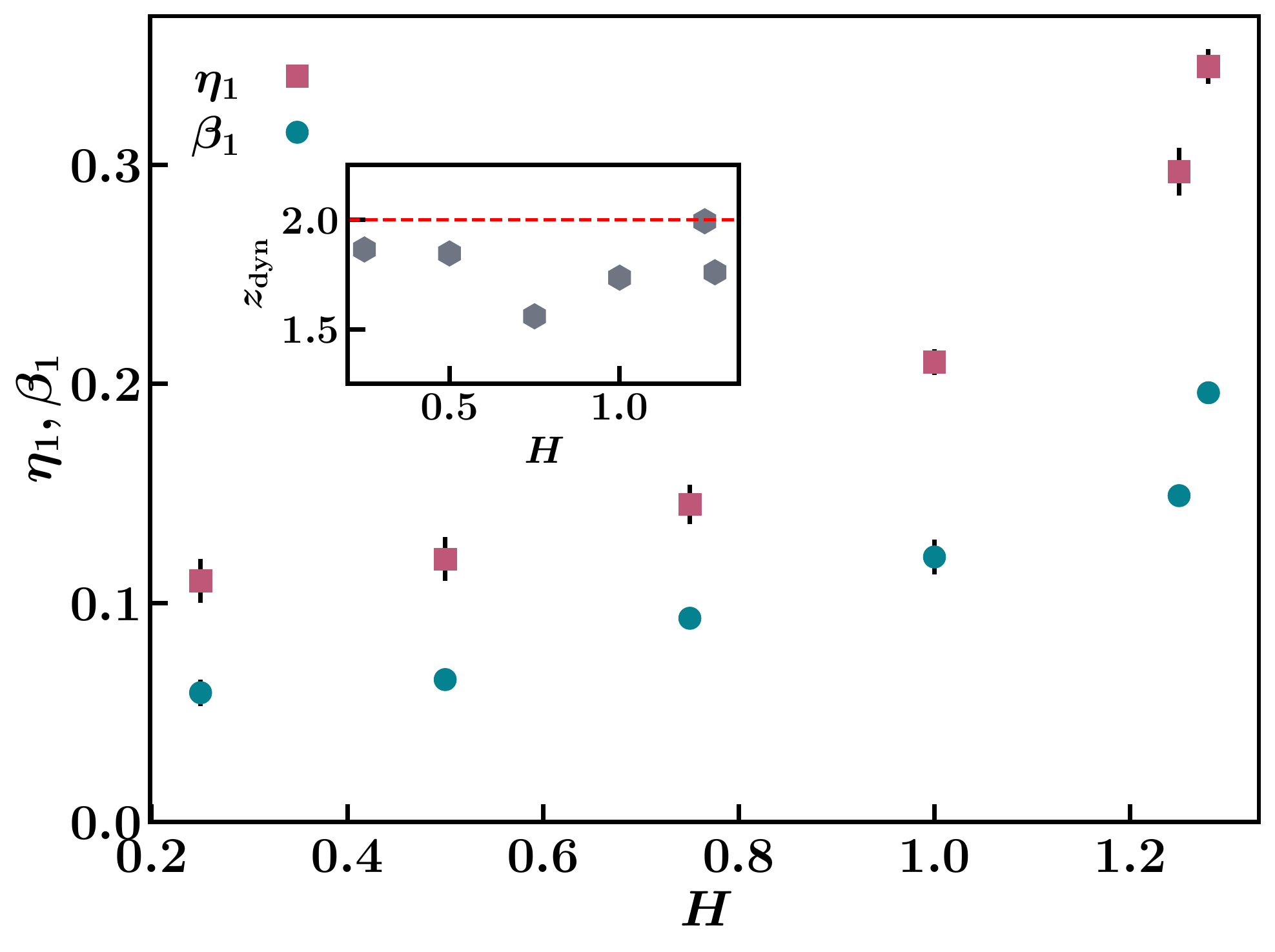}
	\caption{Noise-amplitude dependence of the exponents $\eta_1$ (filled square) and $\beta_1$ (filled circle), at $\tau=1.0$, obtained on a lattice of size $N=100 \times 100$ is shown. The estimated error is shown by the vertical lines. The inset shows that the dynamic exponent (filled hexagon), defined as $z_{\text{dyn}} = {\eta_1}/{\beta_1}$, is $z_{\text{dyn}} \approx 2$.
	}
	\label{fig:exponent_variation_with_H}
\end{figure}

\begin{figure}[]
	\centering
	\includegraphics[scale=0.35]{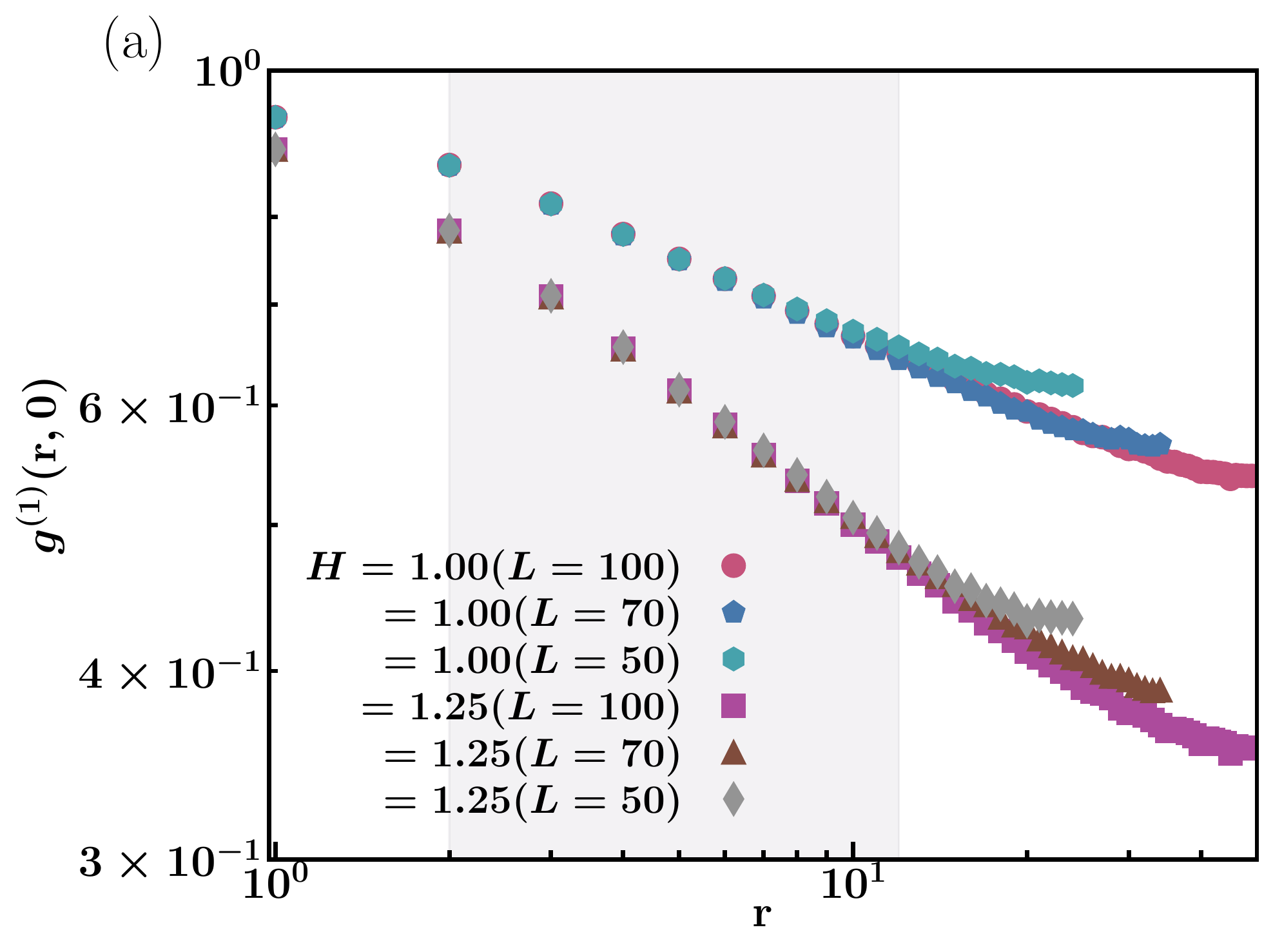}
	\includegraphics[scale=0.365
	]{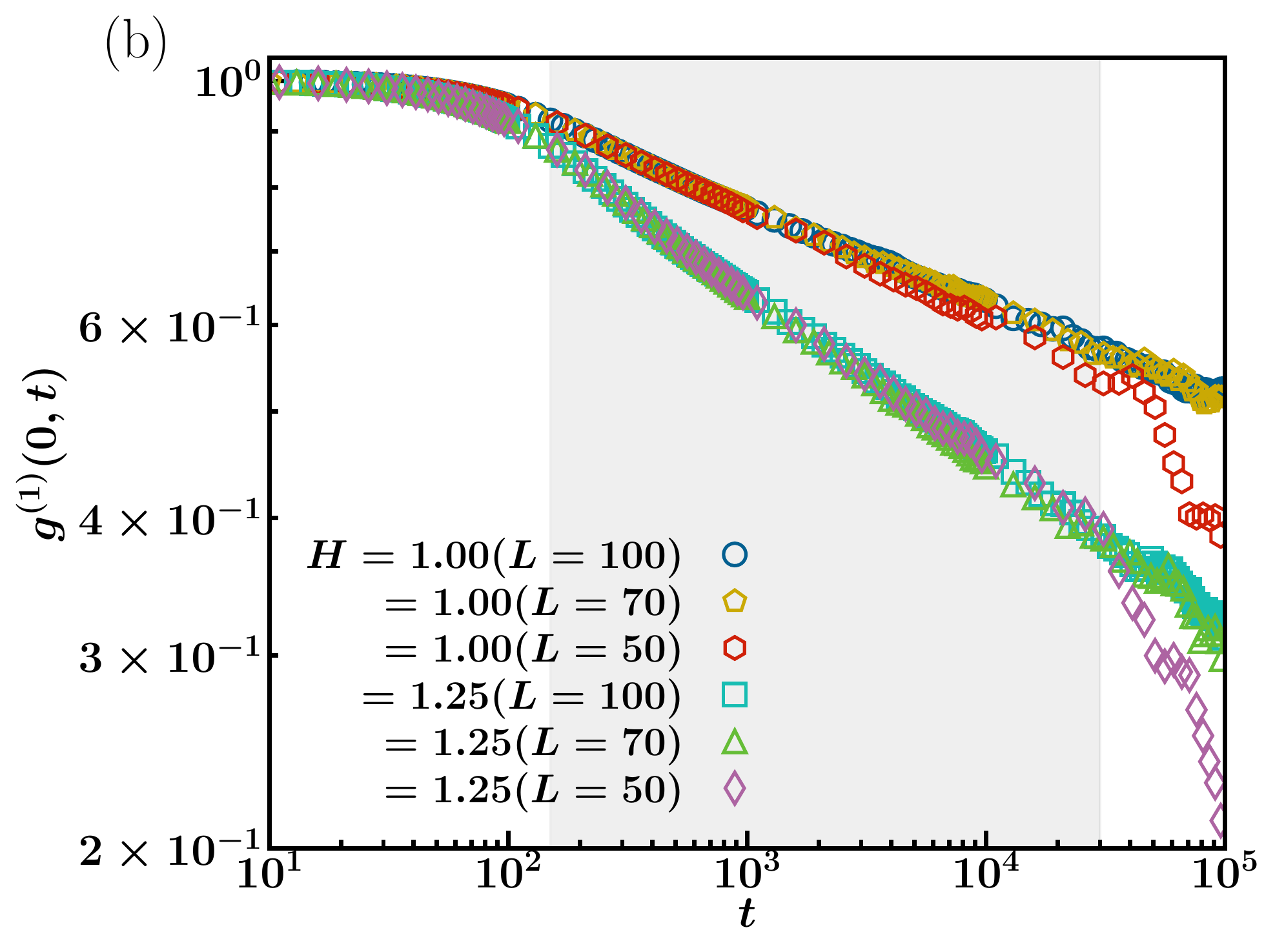}
	\caption{Spatial ($g^{(1)}(r,0)$) and dynamical ($g^{(1)}(0,t)$) correlation, on a log scale, in the nonequilibrium stationary state of the dynamics~(\ref{eq:eom_DN}) on lattice of various sizes $N=L\times L$, with $L=50, 70$ and $100$, computed at two values of dichotomous noise-amplitude, namely, $H=1.0$ and $1.25$ are shown in (a) and (b), respectively. The filled (empty) circle, pentagon and hexagon correspond to spatial (dynamical) correlation for $L=100,~70$ and $50$, respectively at fixed $H=1.0$. Similarly, for fixed $H=1.25$, the spatial (dynamical) correlation for $L=100,~ 70$ and $50$ are denoted by the filled (empty) square, triangle and diamond, respectively. The noise-correlation time is chosen to be $\tau = 1.0$. The distance $r$ is in units of lattice spacing and time $t$ is in integration time steps. As seen from panel (a), the static correlation data for various $L$ collapse in the linear region, implying that the exponents have no considerable system-size dependency. The dynamical correlations, in panel (b), also shows similar behavior. }
	\label{fig:correlators_comparison}
\end{figure}

\begin{figure}[]
	\centering
	\includegraphics[scale=0.375]{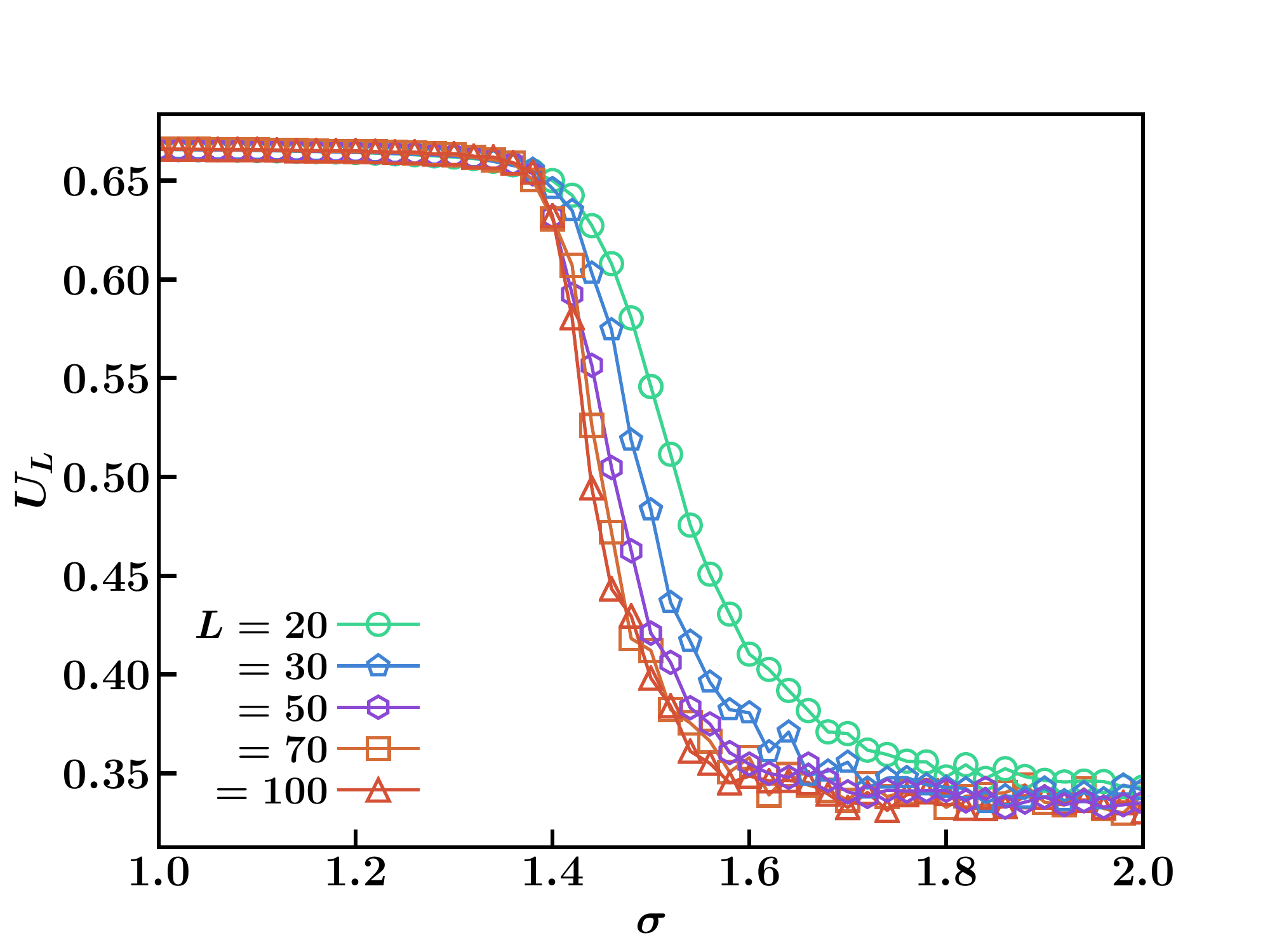}
	\caption{Shown is the variation of Binder cumulant $U_{L}$ with reduced Gaussian white noise strength $\sigma$ in the equilibrium stationary state of the dynamics~(\ref{eq:eom_GWN2}). This corresponds to the Gaussian white noise limit of the dynamics (\ref{eq:eom_DN}), obtained in the simultaneous limit $\tau \to 0$ and $H \to \infty$ keeping $T = H^{2} \tau $ fixed and finite. The parameter $\sigma$ is related to $H$ and $\tau$ as $\sigma = \sqrt{2T/K} = \sqrt{2 H^{2} \tau/K}$. The symbols circle, pentagon, hexagon, square and triangle correspond to $U_{L}$ values for $L=20,~30,~50,~70$ and $100$, respectively. The curves for various $L$ remain collapsed upto $\sigma_c=1.34$, or equivalently, $T_c=0.90$, indicating the existence of a phase with diverging correlation length in the thermodynamic limit in this region. This system thus exhibits $BKT$ transition  between  critically ordered phase at low temperature and a disordered phase at high temperature. The critical temperature obtained is in well agreement with the critical temperature of the $BKT$-transition in the 2D $XY$ model.}
	\label{fig:Binder_cum_GWN}
\end{figure}
\begin{figure}[]
	\centering
	\includegraphics[scale=0.35]{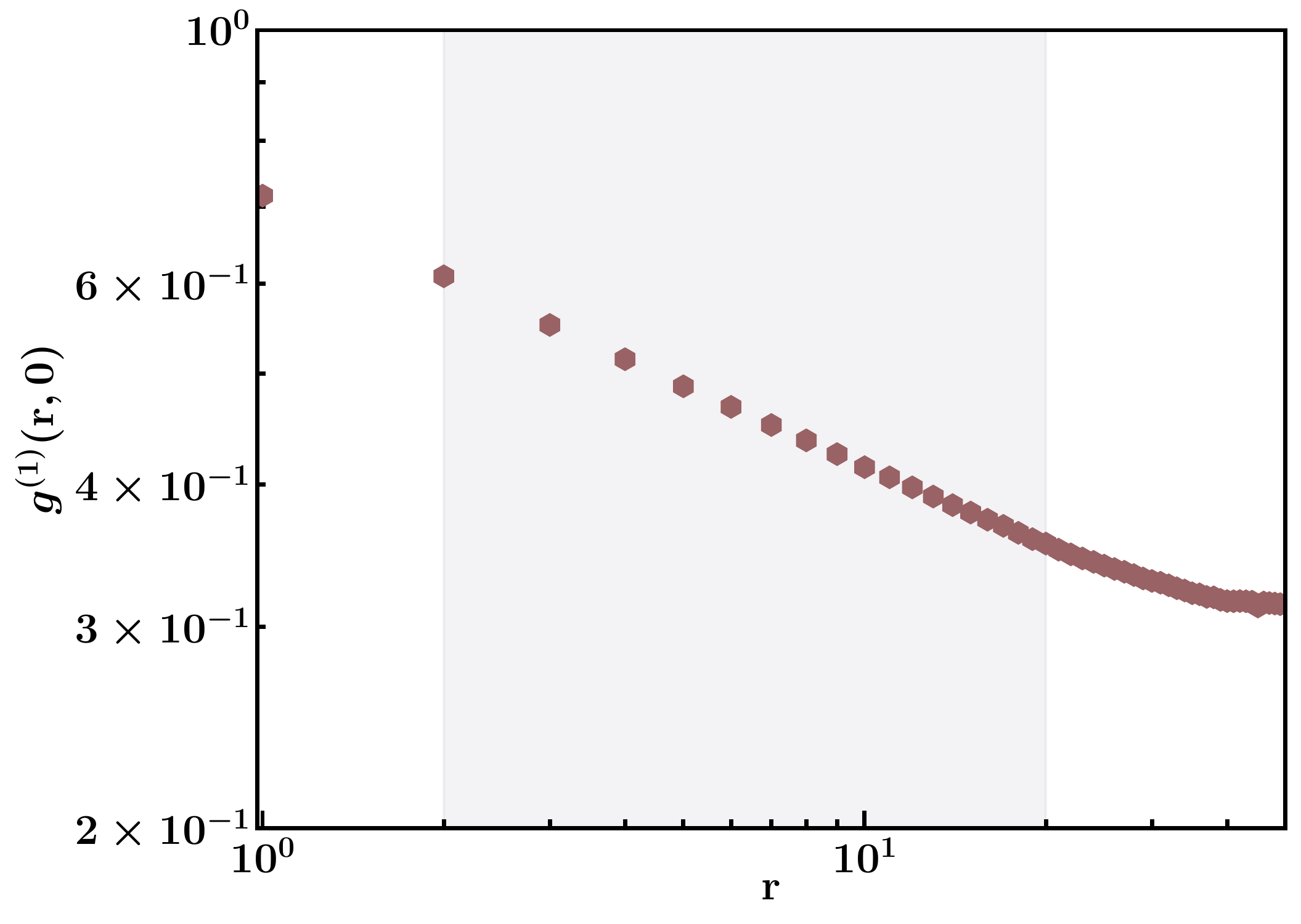}
	\caption{Spatial correlation ($g^{(1)}(r,0)$), on a log scale, in the equilibrium stationary state of the dynamics~(\ref{eq:eom_GWN2}) computed at critical temperature $T_{\text{BKT}} = 0.9$ on a lattice of size $N=100\times 100$, is shown. This corresponds to the Gaussian white noise limit of the dynamics (\ref{eq:eom_DN}), obtained in the simultaneous limit $\tau \to 0$ and $H \to \infty$ keeping $T = H^{2} \tau $ fixed and finite. 
		The distance $r$ is in units of lattice spacing. The power-law exponent obtained is $\eta_1=0.252(4)$, consistent with earlier work.}
	\label{fig:spatial_corr_GWN_T_BKT}
\end{figure}

Fig.~\ref{fig:DN_corr} shows the behavior of $g^{(1)}(r,0)$ and $g^{(1)}(0,t)$, on a log scale, at various noise amplitudes $H$ for $\tau =1.0$ on a particular lattice of size $N=100 \times 100$.
We note that we look only at large distance and long time behavior of the spatial and dynamical correlations, respectively.
But, there is pronounced finite-size effect at large scales in the behavior of spatial correlations, and fluctuations at long times in dynamical correlations arising from stochasticity in the system. Thus, to characterize the correlation behavior, we choose a suitable spatial (temporal) window as shown by shaded region in Fig.~\ref{fig:DN_corr}.

For a fixed $\tau$, as the noise amplitude $H$ increases, the spatial correlation changes its behavior as seen from Fig.~\ref{fig:DN_corr}(a). Upto a critical $H$-value, there is an algebraic decay of the correlation, $g^{(1)}(r,0) \sim r^{-\eta_1}$; beyond which the correlation falls off exponentially fast, $g^{(1)}(r,0) \sim  e^{-r/r_s}$. The dynamical correlation function, plotted on a log scale in Fig.~\ref{fig:DN_corr}(b), also shows similar behavior. It decays algebraically upto the same critical $H$-value,  $g^{(1)}(0,t) \sim t^{-\beta_1}$, and beyond that, the decay is exponential, $g^{(1)}(0,t) \sim e^{-t/t_d}$. The critical noise amplitude $H_c$ at which there is a crossover from algebraic to exponential decay matches with that obtained from the behavior of $U_{L}$.

The critical exponents $\eta_1$ and $\beta_1$ for various $H$-values and $\tau=1.0$, extracted from the power law fit in the linear regime of the static and dynamic correlators, on a lattice of size $N=100 \times 100$ are shown in Fig.~\ref{fig:exponent_variation_with_H}. In the critically ordered phase, critical indices vary continuously with noise amplitude $H$, yielding the ratio $ {\eta_1}/{\beta_1} \approx 2$. This indicates that the dynamical exponent $z_{\text{dyn}}$, defined as $ z_{\text{dyn}} = {\eta_1}/{\beta_1}$, is approximately $2$. This phase transition from low-$H$ critically ordered phase with algebraic decay of correlation to high-$H$ disordered phase with an exponential decay of correlation is analogous to the Berezinskii-Kosterlitz-Thouless ($BKT$) transition as observed in the 2D $XY$ model \cite{kosterlitz1973ordering,kosterlitz1974critical}. We note that the maximum value of the exponent $\eta_1 \approx 0.35$ exceeds the equilibrium upper bound limit i.e. $0.25$ \cite{kosterlitz1973ordering,kosterlitz1974critical}. This may be due to either finite-size effects or the nonequilibrium nature of the dynamics, or both. To investigate the finite-size effect, for the same noise correlation time $\tau=1.0$, we compute the spatial correlation functions at $H=1.0$ and $1.25$ on lattice of various sizes, $N=50 \times 50$, $70 \times 70$, and $100 \times 100$. These are shown in Fig.~\ref{fig:correlators_comparison}. As observed from Fig.~\ref{fig:correlators_comparison}(a), the spatial correlation data for various $L$ collapse in the linear regime suggesting that there is no appreciable dependence of the exponents on system-size. Dynamical correlation, shown on a log scale in Fig.~\ref{fig:correlators_comparison}(b), also shows similar behavior.


\paragraph*{Dynamics in presence of Gaussian white noise:}
	To this end, we numerically study the dynamics~(\ref{eq:eom_DN1}) in the equilibrium limit, treating $\zeta(t)$ as a Gaussian white noise.
	The evolution equation now reads as
	\begin{equation}
		\frac{{\rm d} \theta_{i}}{\rm dt} =  K  \sum_{j \in nn_{i}}  \sin(\theta_{j}- \theta_{i}) + \sqrt{2T} \zeta_{i}(t),
		\label{eq:eom_GWN}
	\end{equation}
	where the term $\zeta_{i}(t)$ is a Gaussian white noise characterized by
	\begin{equation}
		\langle \zeta_{i}(t) \rangle = 0 ~\text{and}~ \langle \zeta_{i}(t)\zeta_{j}({t'}) \rangle = \delta_{ij} \delta(t - {t}^{\prime}).
	\end{equation}
	Here $\langle \cdot \rangle$ denotes averaging over noise realizations, and $T$ is the noise strength which represents essentially the temperature of the system. We further implement for $K \neq 0$ the following transformation
	\begin{equation}
		\nonumber
		{t} \to Kt, ~ {\sigma} \to  \sqrt{2T/K}  ~ \mbox{and}~ {\zeta}_{i}({t}) \to {\zeta}_{i}(t)/K,  
	\end{equation}
	to reduce the governing dynamics (\ref{eq:eom_GWN}) to a dimensionless form as follows:
	\begin{equation}
		\frac{{\rm d} \theta_{i}}{\rm d\tilde{t}} =  \sum_{j \in nn_{i}}  \sin(\theta_{j}- \theta_{i}) + {\sigma} {\zeta}_{i}({t}).
		\label{eq:eom_GWN2}
	\end{equation}
Statistical quantities are measured in the equilibrium stationary state attained at long time by numerically integrating the dynamics (\ref{eq:eom_GWN2}) employing Euler-Maruyama algorithm with integration time step $dt=0.01$ \footnote{Note that during integration with white noise, the random number (which acts as white noise force) was sampled from a stationary distribution and thus stationarity of only the Kuramoto system was checked in this case.}. Periodic boundary conditions are applied.
	
The system (\ref{eq:eom_GWN2}) exhibits $BKT$ transition as observed from the behavior of $U_L$ with reduced noise strength $\sigma$ for various values of $L$, shown in Fig.~\ref{fig:Binder_cum_GWN}. We obtain the critical reduced noise strength, as the point upto which the curves of $U_L$ for various $L$ stay collapsed, $\sigma_c = 1.34$. This is equivalent to critical temperature $T_c=0.90$, which is in well agreement with the critical temperature of the $BKT$ transition in the 2D $XY$ model obtained via Monte Carlo simulation \cite{tobochnik1979monte,fernandez1986critical}. The spatial correlation $g^{(1)}(r,0)$, shown on a log scale in Fig.~\ref{fig:spatial_corr_GWN_T_BKT}, computed at critical temperature $T_{\text{BKT}}=0.9$ yields the power law exponent $\eta_1=0.252(4)$, which is also consistent with the previous work \cite{kosterlitz1973ordering,kosterlitz1974critical}.

Based on the foregoing discussion, we believe that the exponent $\eta_1 = 0.35$ is a signature of nonequilibrium nature of the transition. Such value of the exponent has been reported in earlier works on 2D planar model \cite{luther1977critical, nelson1977universal} and recently in driven-dissipative condensates \cite{dagvadorj2015nonequilibrium,comaron2021non}.
Next, we repeat the same study for various values of noise correlation time $\tau$, and for each $\tau$, we calculate the critical noise amplitude $H_c$ from the $U_L$ against $H$ curves for various $L$, and this was further confirmed from the behavior of correlation functions. Having obtained these, we construct the phase-diagram in the relevant parameter space for the system, which will be discussed in the next section.

\subsection{Phase Diagram}

\begin{figure}[]
	\centering
	\includegraphics[scale=0.35]{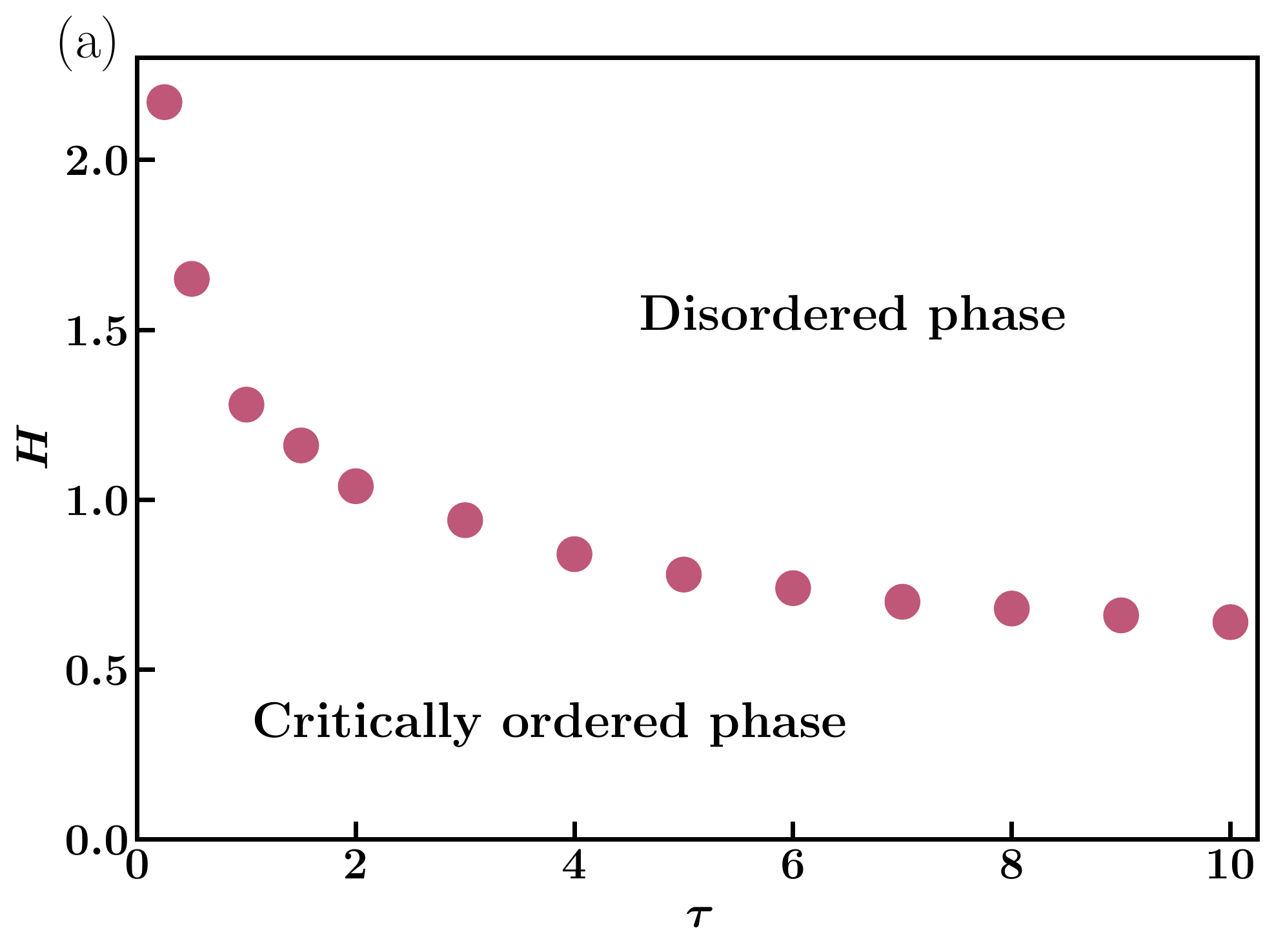}
	\includegraphics[scale=0.35]{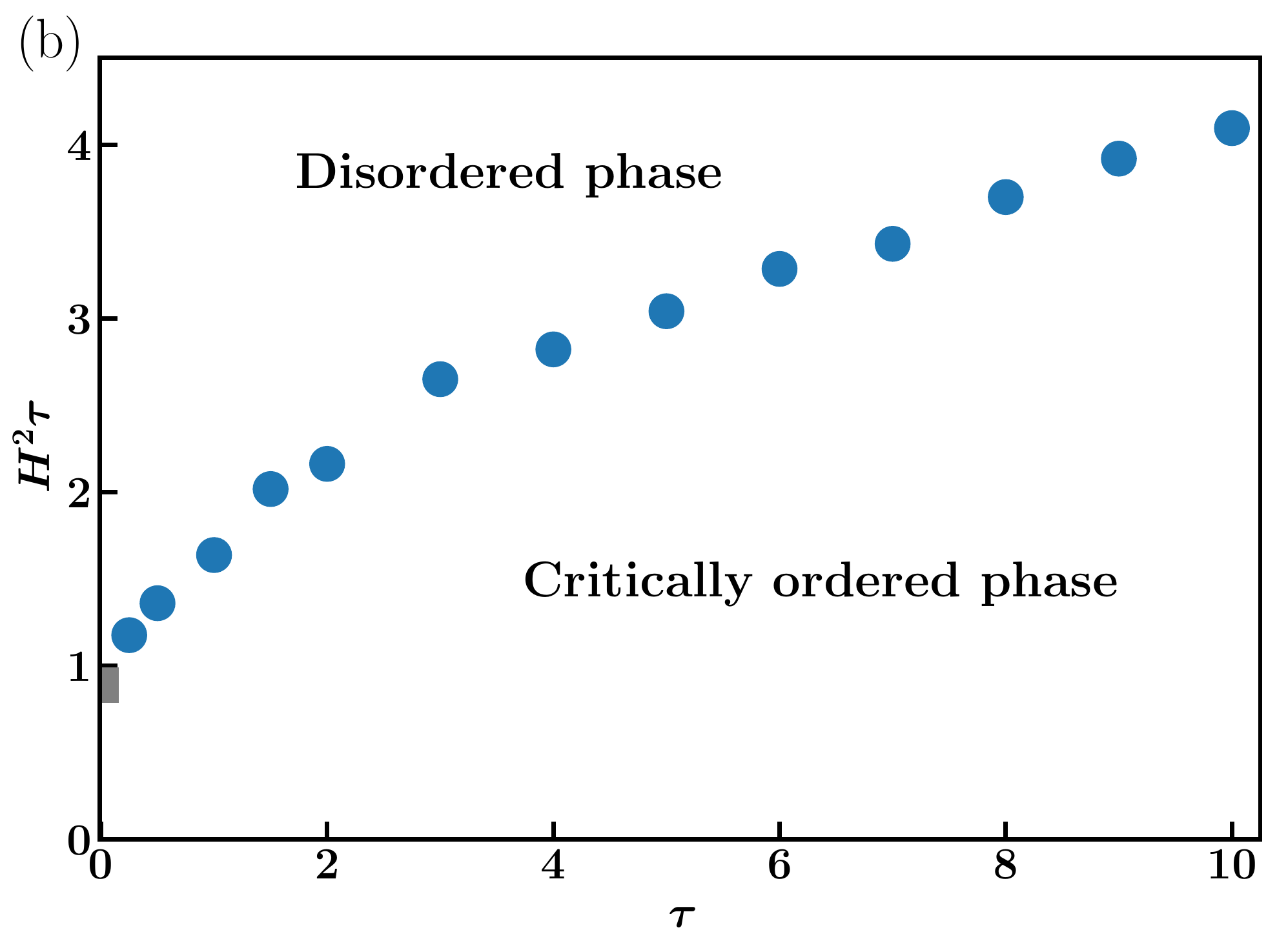}
	\caption{The complete, nonequilibrium stationary-state phase diagram of the dynamics~(\ref{eq:eom_DN}) is shown in the
		($H-\tau$) (a) and ($H^{2}\tau-\tau$)(b) plane. Both the panels show the nonequilibrium $BKT$-like transition line (filled circles), separating the two distinct phases: critically ordered (quasi-ordered) phase characterized by algebraic decay of correlation at low noise-amplitude and disordered phase with exponential decay of the correlation at high noise-amplitude. Panel (b) shows additionally the critical temperature of the equilibrium $BKT$ transition, given by $T_{BKT}=H^2\tau \approx 0.9$, shown by a grey square, as recovered from a study of a suitable limiting case of the dynamics~(\ref{eq:eom_DN}). In the limit $\tau \to 0$, the line of transition in ($H^{2}\tau-\tau$) plane tends to hit the $y$-axis at that equilibrium critical temperature, showing consistency of our work.
	}
	\label{fig:DN_parameter_space}
\end{figure}

The complete, nonequilibrium stationary-state phase diagram of the dynamics~(\ref{eq:eom_DN}) in the
($H-\tau$) and ($H^2\tau-\tau$) plane is shown in Fig.~\ref{fig:DN_parameter_space}(a) and Fig.~\ref{fig:DN_parameter_space}(b), respectively. It is clear from the phase diagram that for all finite correlation times, there exist two distinct phases: critically ordered (quasi-ordered) phase characterized by algebraic decay of correlation at low noise amplitude and a disordered phase with exponential decay of the correlation at high noise amplitude, and the system exhibits nonequilibrium $BKT$-like transition between them as one tunes the parameter $H$ from low to high value. Interestingly, the critical noise amplitude $H_c$ decreases with an increase in noise correlation time $\tau$. We will provide a qualitative argument in the following section.

To visualize the white noise limit, we replot the phase diagram with change of parameters, namely, in ($H^2\tau-\tau$) plane as shown in Fig.~\ref{fig:DN_parameter_space}(b). The white noise limit is achieved in the limit $\tau \to 0$ keeping $T=H^{2}\tau$ finite. The critical temperature of the equilibrium $BKT$ transition, given by $T_{BKT}=H^2\tau \approx 0.9$ is shown by a grey square in Fig.~\ref{fig:DN_parameter_space}(b).
In the limit $\tau \to 0$ keeping $T=H^{2}\tau$ finite, the line of transition tends to hit the $y$-axis close to that equilibrium critical temperature. This shows the consistency of our work.

The mechanism behind this nonequilibrium phase transition is believed to be the same as the equilibrium one. There are always vortices and spin-waves present in the nonequilibrium stationary state of the system. Here the oscillators are treated as spins with the direction indicating the phase of the oscillator. For a fixed $\tau$, in the region $H \leq H_c$, the vortices are bound in pairs with total vorticity zero, and spin-wave excitations are the dominant ones. The vortices and antivortices annihilate and thus the system is free of defects. The spin-wave excitations are responsible for destroying long range order in the system. But above $H_c$, the vortices become unbound and they are now free to move to the surface, thereby causing a phase transition.

\begin{figure}[]
	\centering
	\includegraphics[scale=0.2]{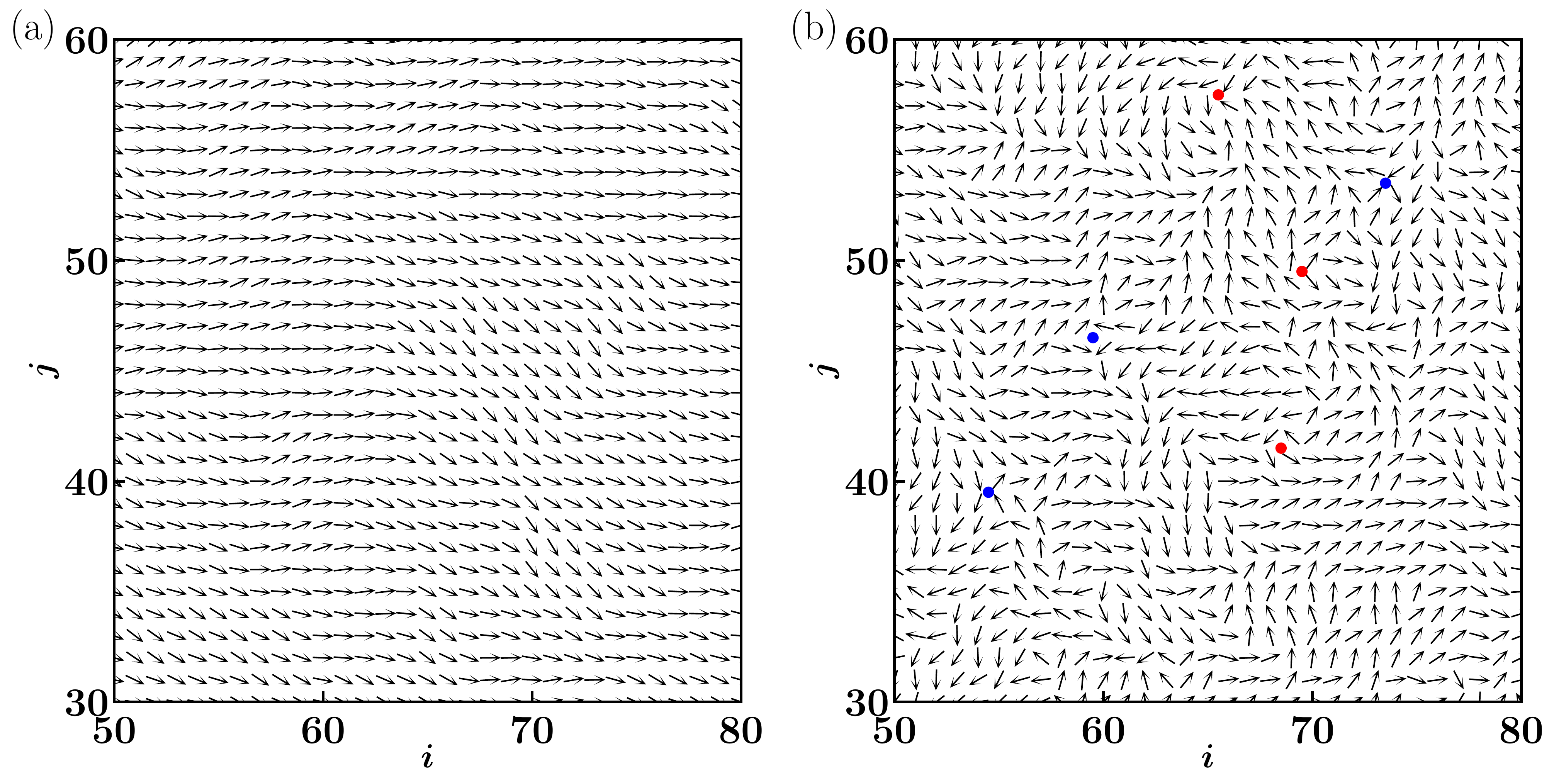}
	\caption{Shown is the orientation of the oscillators' phase on a portion of a lattice of size $N=100 \times 100$, in the nonequilibrium stationary state of the dynamics~(\ref{eq:eom_DN}), at two values of dichotomous noise amplitude, namely, $H=1.0$(a) and $1.5$(b). The topological configurations in (a) and (b) correspond to the time snapshots (b) and (d) in Fig.~\ref{fig:DN_snapshots_all}, respectively. Oscillators are represented by the indices ($i,j$).
	Each arrow represents one oscillator pointing in a direction according to the phase of the oscillator 
	The noise-correlation time is chosen to be $\tau =1.0$. While panel (a) shows spin-waves only, panel (b) displays unbound vortices and anti-vortices (``Topological defects'') which are free to proliferate. Few vortices (denoted by red dots) and antivortices (denoted by blue dots) are shown.
	}
	\label{fig:DN_vortices}
\end{figure}

Fig.~\ref{fig:DN_vortices} depicts this scenario. Here the orientation of the oscillators' phases (topological configurations) for $\tau=1.0$, at two values of $H$, namely, $H=1.0 (H < H_c)$ and $H=1.5 (H > H_c)$, from a portion of a lattice of size $N=100 \times 100$, is shown in (a) and (b), respectively. Fig.~\ref{fig:DN_vortices}(a) displays spin waves only, and absence of defects in the phase field of the oscillators, 
whereas Fig.~\ref{fig:DN_vortices}(b) shows the presence of unbound vortices and anti-vortices (topological defects) which are free to proliferate. Few of the vortices and anti-vortices are shown by red and blue dots, respectively. 

\begin{figure}
	\centering
	\includegraphics[scale=0.25]{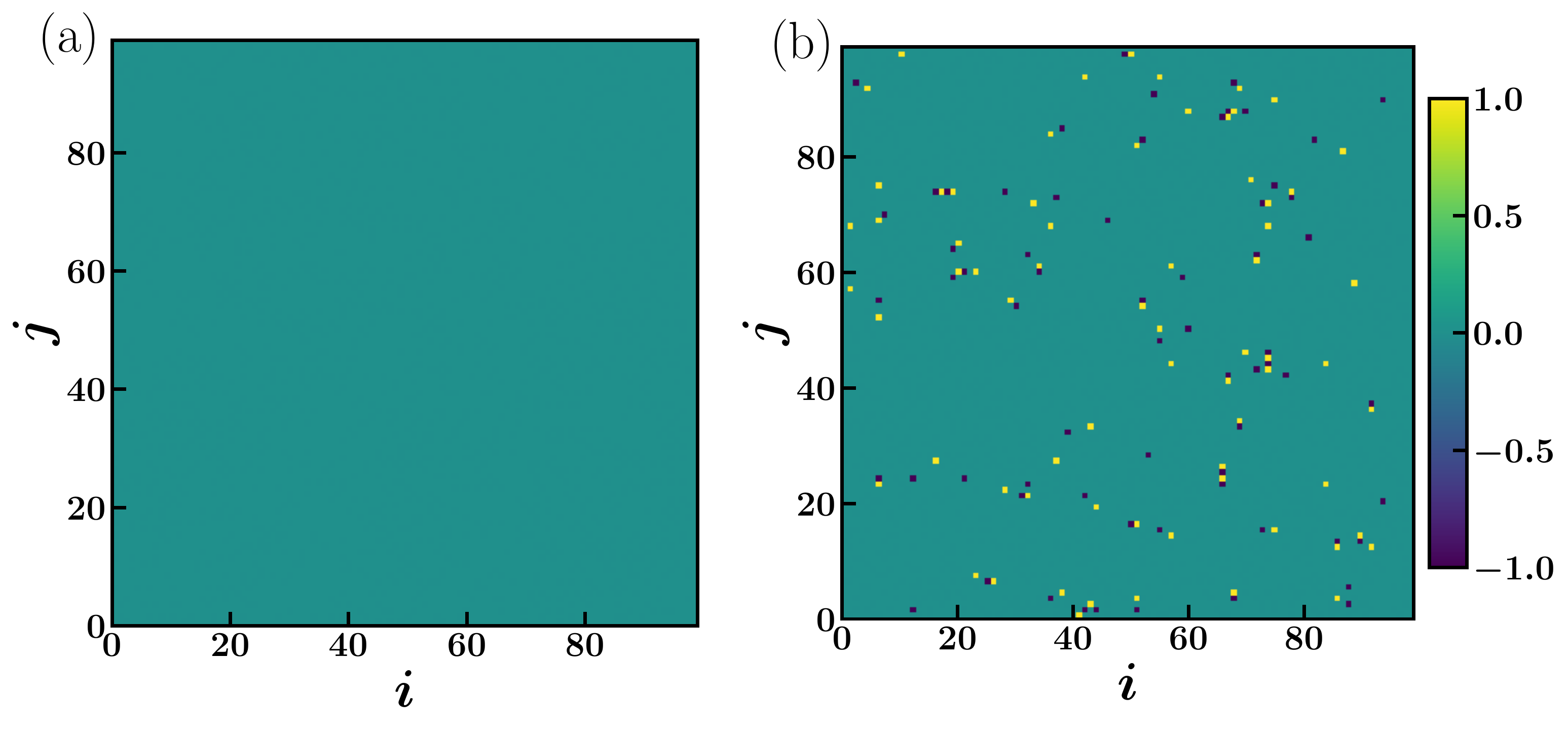}
	\caption{Shown are snapshots of the vorticity field in the nonequilibrium stationary state of the dynamics~(\ref{eq:eom_DN}) on a lattice of size $N=100 \times 100$, at two values of dichotomous noise amplitude, namely, $H=1.0$(a) $(H < H_c)$~\text{and}~$1.5$(b) $(H > H_c)$. These vorticity field snapshots correspond to the phase snapshots (b) and (d) in Fig.~\ref{fig:DN_snapshots_all}, respectively. Oscillators are represented by the indices ($i,j$). Each pixel represents a unit of four oscillators and the color indicates its vorticity, where $+1$ (vortex of unit strength) is represented by yellow while $-1$ (antivortex of unit strength) is by black. The noise-correlation time is chosen to be $\tau =1.0$. While panel (a) shows a phase which is free of vortices, panel (b) displays unbound vortices and anti-vortices which are free to proliferate.}
	\label{fig:DN_vorticity_snapshot}
\end{figure}

\begin{figure}
	\centering
	\includegraphics[scale=0.35]{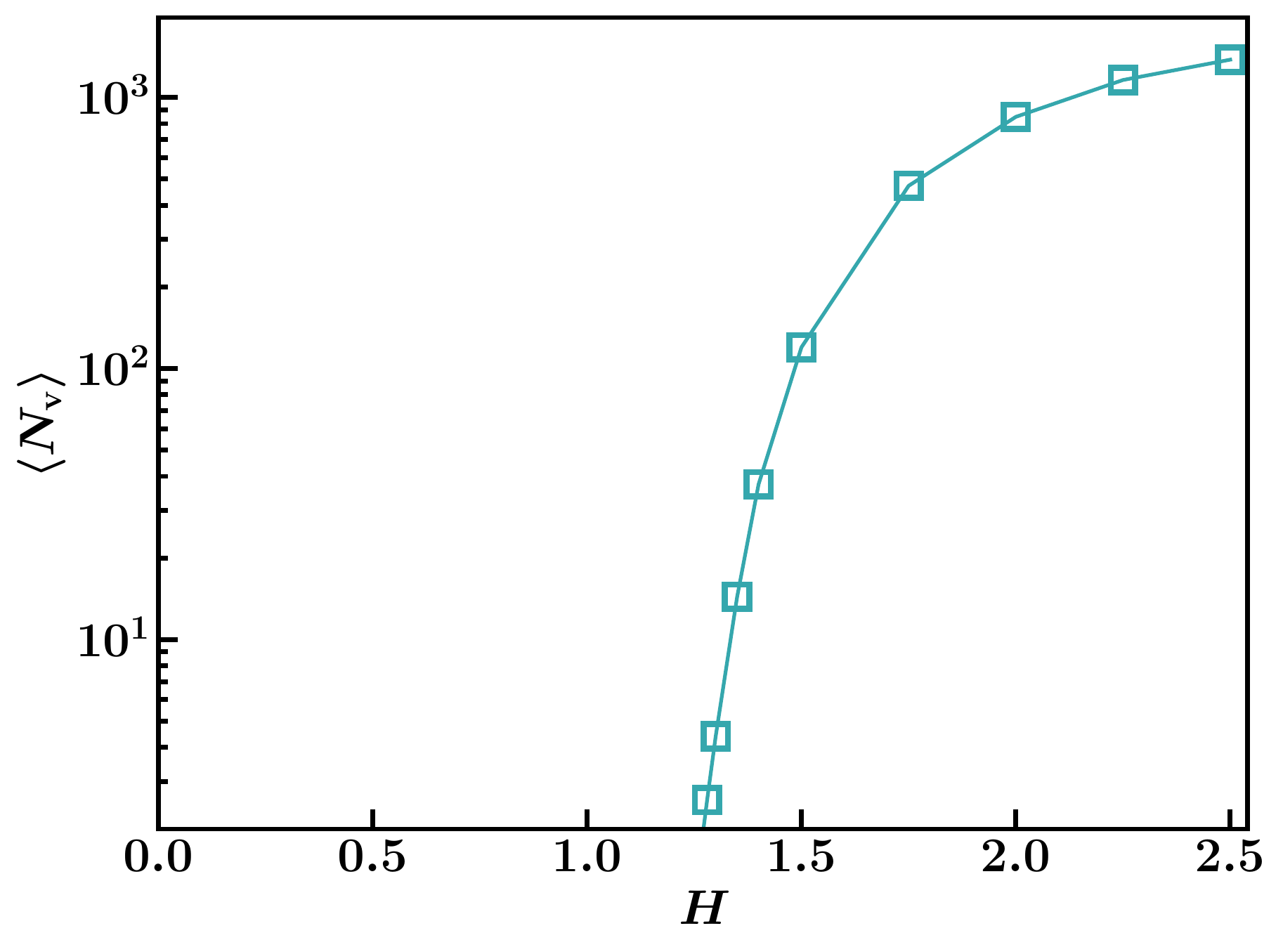}
	\caption{Variation of average number of topological defects (vortices and antivortices) $N_{\rm v}$ as a function of dichotomous noise amplitude $H$, on a semilog scale, in the nonequilibrium stationary state of the dynamics~(\ref{eq:eom_DN}) computed on a lattice of size $N=100\times 100$, is shown. The noise-correlation time is chosen to be $\tau =1.0$. The system is free of defects in the region $H < H_{c}$, whereas near the critical point the vortices starts unbinding and $\langle N_{\rm v} \rangle$ shows a sharp rise right beyond the critical value $H_c$.}
	\label{fig:DN_total_vortices}
\end{figure}

To visualize the picture more clearly, we identify the vortices and compute their strength, i.e. vorticity numerically in the following way\footnote{	
	The vorticity, in the continuum limit, is defined as follows:
	\begin{equation}
		\frac{1}{2\pi}\oint \nabla\theta(\hat{r} ,t).d\hat{l} = \pm  n
	\end{equation}
	where $d\hat{l}$ is the integration path enclosing the defect and $n$ is called the topological charge or vorticity. The defect is said to have charge $n = +1$ if the integral is $ +2 \pi$ (for a vortex), $n=-1$ if it is $ -2 \pi$ (for an antivortex) and $n=0$  if it is zero (for no vortex).}.
For each oscillator on the lattice, say $(i,j)$-th oscillator, we consider a plaquette of four oscillators, namely, $(i,j)$-th, $(i+1,j)$-th, $(i+1,j+1)$-th and $(i,j+1)$-th and compute the lattice curl of the phase gradient around this  plaquette (unit cell of four oscillators). The curl is equal to the sum of directed phase differences with modulo $2\pi$. The vorticity thus computed is assigned to the $(i,j)$-th oscillator. 
Figure~{\ref{fig:DN_vorticity_snapshot}} shows snapshots of vorticity field for $\tau = 1.0$ and at two values of $H$, namely, $H=1.0$ and $1.5$ in (a) and (b) respectively. These snapshots correspond to the phase snapshots (b) and (d) in Fig.~\ref{fig:DN_snapshots_all}, respectively. Figure~{\ref{fig:DN_vorticity_snapshot}}(a) corresponds to the critically ordered phase $(H < H_{c})$ and thus free of defects. On the other hand, Fig.~{\ref{fig:DN_vorticity_snapshot}}(b) displays unbound vortices and antivortices characteristics of desynchronized phase. To verify this picture quantitatively, we further calculate the total number of defects (vortices and antivortices) $ N_{\rm v}$ at various $H$-values keeping $\tau$ fixed and average this quantity over $200$ realizations. The quantity $\langle N_{\rm v} \rangle$ is zero in the region $H < H_c$. Near $H_c$, $\langle N_{\rm v} \rangle$ takes small value which suggests initiation of unbinding of vortices-antivortices. But, it sharply rises right beyond the critical point suggesting the plasma of free vortices and antivortices. This is evident from Fig.~{\ref{fig:DN_total_vortices}}.

\subsection{Dynamics of a single Topological defect}

\begin{figure}
	\centering
	\includegraphics[scale=0.37]{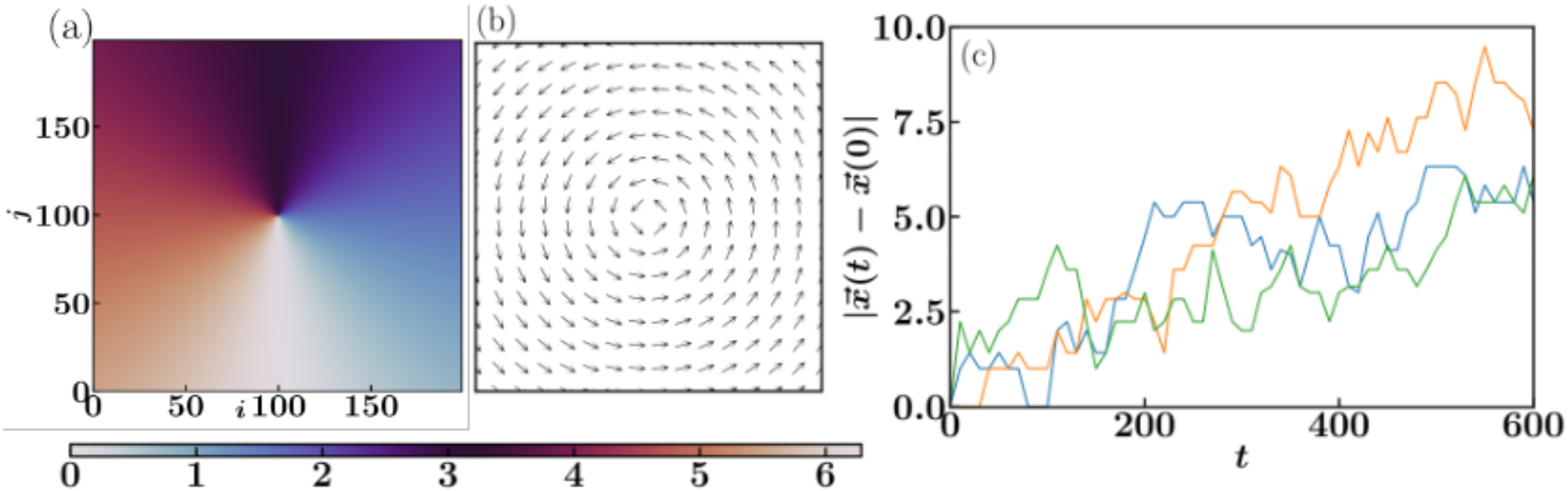}
	\caption{(a): Snapshot showing initial configuration of the oscillators' phase hosting a single vortex at the center of a lattice of size $N= 200 \times 200$. Oscillators are represented by the indices ($i,j$). Each pixel represents one oscillator and the color indicates its phase as denoted by the color bar.
	(b): Topological configuration of the defect at and around the center of the lattice is shown. Each arrow represents one oscillator pointing in a direction according to its phase.
	(c): Three typical realizations of displacement $|\vec{x}(t) - \vec{x}(0)|$ made by the defect on the lattice. The noise amplitude and correlation time are chosen to be $H=0.8$ and $\tau = 0.5$, respectively.
	}
	\label{fig:DN_single_vortex}
\end{figure}

\begin{figure}
	\centering
	\includegraphics[scale=0.20]{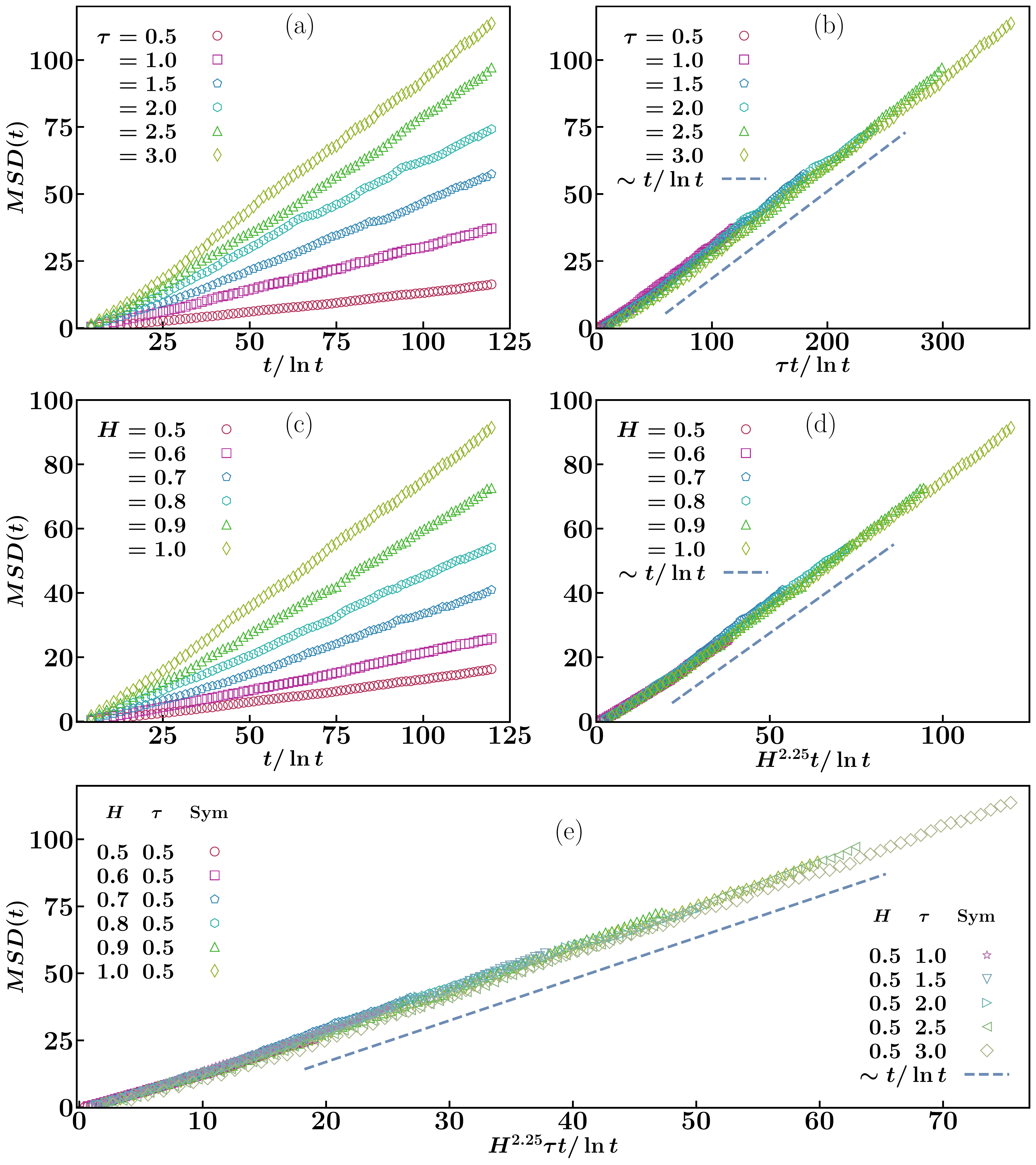}
	\caption{(a): Mean-squared displacement, MSD made by a single topological defect (vortex) on a lattice of size $N= 200 \times 200$, for a fixed noise amplitude $H=0.5$ and various noise correlation time $\tau$. The symbols circle, square, pentagon, hexagon, triangle and diamond correspond to MSD for $\tau=0.5,~1.0,~1.5,~2.0,~2.5$ and $3.0$, respectively. The curves for various $\tau$ are plotted on a linear scale with $t/{\ln t}$. They show a linear behavior.
	(b): The MSD data for various $\tau$, denoted by the same symbols as in panel (a), are plotted on a linear scale with $\tau t/{\ln t}$. They collapse on a single curve suggesting that the coefficient ${\cal{D}} \propto \tau$ for a fixed $H$. 
	(c): MSD for a fixed noise correlation time $\tau=0.5$ and various noise amplitudes $H$ are displayed. The symbols circle, square, pentagon, hexagon, triangle and diamond correspond to MSD for $H=0.5,~0.6,~0.7,~0.8,~0.9$ and $1.0$, respectively. The curves for various $H$ show a linear behavior with $t/{\ln t}$.
	(d): The data for various $H$, denoted by the same symbols as in panel (c), are plotted on a linear scale with $H^{\alpha} t/{\ln t}$ with $\alpha = 2.25$. A good collapse of the data suggests that the coefficient ${\cal{D}} \propto H^{\alpha}$ for a fixed $\tau$.
	(e): The data for various $H$ and $\tau$  are plotted on a linear scale with $H^{\alpha} \tau t/{\ln t}$. The symbols circle, square, pentagon, hexagon, triangle-up and thin-diamond correspond to MSD for $H=0.5,~0.6,~0.7,~0.8,~0.9$ and $1.0$, respectively, at a fixed $\tau=0.5$. The symbols star, triangle-down, triangle-right, triangle-left and thick-diamond correspond to MSD for $\tau=1.0,~1.5,~2.0,~2.5$ and $3.0$, respectively, at a fixed $H=0.5$. A good collapse is evident suggesting ${\cal{D}} \propto {H^{\alpha} \tau}$. The dotted line in panels (b), (d) and (e) indicates linear behavior of MSD with $t/{\ln t}$ which confirms the anomalous diffusive behavior.
	}
	\label{fig:DN_MSD}
\end{figure}

In this section we try to understand the phase diagram, the interplay between  critical $H$ and $\tau$-values qualitatively. As we observed that the phase transition is caused due to unbinding of topological defects, we turn our discussion to defects only. We investigate the dynamics of a single defect under the influence of dichotomous noise. We prepare the system initially in such a way that the whole lattice hosts a single defect at its center only and let the system evolve. After we initialized the oscillators phases this way, we numerically integrate the dynamics~(\ref{eq:eom_DN}) with fourth order Runge-Kutta method with integration time step $dt=0.01$ imposing open boundary conditions. During its evolution, we ensure that the motion of the defect does not get affected by the boundary and no further defects are created. We implement the same setting as in \cite{nogawa2009abnormal}. Keeping this in mind, we perform the study of defect dynamics on a larger lattice, of size $N=200 \times 200$. Figure~{\ref{fig:DN_single_vortex}}(a) shows a snapshot of the oscillators' initial phase showing that the lattice hosts a single defect (vortex of unit strength) at its center. The corresponding topological configuration of the defect around the center of the lattice is shown in Fig.~{\ref{fig:DN_single_vortex}}(b). Figure.~{\ref{fig:DN_single_vortex}}(c) shows three typical realizations of displacement $|\vec{x}(t) - \vec{x}(0)|$ performed by the defect on the lattice for $H=0.8$ and $\tau = 0.5$. Here $\vec{x}(t) = (x(t), y(t))$ is the position of the defect on the lattice at time $t$.

The mean-squared displacement (MSD) of the vortex is evaluated for a fixed $H$ and $\tau$. All the data for MSD presented here are obtained by averaging over $2000$ realizations. We observe that the dynamics of the defect does not obey normal diffusion displaying linear growth of MSD in time $t$. Instead, a logarithmic correction to the normal diffusion is observed and the MSD scales linearly with $t/{\ln t}$ which suggests that this diffusive motion is slower than normal diffusion. This is shown in Fig.~{\ref{fig:DN_MSD}}. This kind of anomalous diffusive behavior is similar to that in presence of Gaussian white noise \cite{nogawa2009abnormal}.  We thus write MSD, $\langle |\vec{x}(t) - \vec{x}(0)|^{2} \rangle = 4 {\cal{D}} t/{\ln t}$, where $\cal{D}$ is the pseudo-diffusion coefficient.

We now study the effect of noise amplitude $H$ and correlation time $\tau$ on MSD results, to be more precise, on the pseudo-diffusion coefficient $\cal{D}$. We calculate the MSD for a fixed $H=0.5$, and various $\tau$. These are shown in Fig.~{\ref{fig:DN_MSD}}(a). We observe that for a fixed $H$, the coefficient ${\cal{D}} \propto \tau$, as a result of which the MSD data for various $\tau$, when plotted as a function of $\tau t/{\ln t}$, collapse on a single curve as evident from Fig.~{\ref{fig:DN_MSD}}(b). The dotted line in blue indicates the linear relationship of MSD with $ t/{\ln t}$. On the other hand, for a fixed $\tau$, the MSDs for various $H$ are shown in Fig.~{\ref{fig:DN_MSD}}(c). We observe in this case ${\cal{D}} \propto H^{\alpha}$ with $\alpha \approx 2.25$. This results in collapsing the data for various $H$ when plotted on a linear scale with $H^{\alpha} t/{\ln t}$, as shown in Fig.~{\ref{fig:DN_MSD}}(d). Thus, for dichotomous noise of amplitude $H$ and correlation time $\tau$, the coefficient follows ${\cal{D}} \propto  H^{\alpha}\tau$. As a result, a good scaling collapse of the MSD data for various $H$ and $\tau$ is evident in Fig.~{\ref{fig:DN_MSD}}(e), plotted on a linear scale as a function of $H^{\alpha} \tau t/{\ln t}$. The dotted line in blue confirms the linear relationship of MSD with $ t/{\ln t}$ suggesting the anomalous diffusive behavior. 

We thus observe that for a fixed $H$, an increase in $\tau$ increases the coefficient $\cal{D}$ or equivalently the mobility of the defect in the system. We further examine in our system and found that there is creation of defects at high $\tau$ with rate of creation increasing with an increase in $\tau$. We believe that due to increased creation rate of defects, the vortices become unbound at lower $H$-value, creating a state of free vortices-antivortices causing a phase transition. Consequently, an increase in $\tau$ decreases the critical $H$-value of the $BKT$-like transition. We note that the dynamics of a single isolated defect is very different from the collective dynamics of several defects interacting with each other via an effective potential. But we believe the knowledge of dependence of mobility of a single defect on $H$ and $\tau$ helps us understand the behavior of pair of defects in the system, and we hope it might be helpful in the study of many defects too. For instance, for a pair of defects, vortices with higher mobility are expected to have more tendency to become unbound. However, a complete confirmation of this by studying vortex-vortex interaction and the dynamics of bound pair of defects (vortex-antivortex) with $H$ and $\tau$ requires an independent study, and thus is beyond the scope of our present work.



\section{Conclusions}
\label{sec:conclusions}
In summary, we have systematically explored the impact of dichotomous noise on synchronization in the locally coupled Kuramoto model with identical natural frequencies arranged on the sites of a 2D periodic square lattice. We show that the resulting dynamics~(\ref{eq:eom_DN}) exhibits a nonequilibrium $BKT$-like transition between a phase with quasi long-range order characterized by algebraic decay of correlation at low noise amplitude and a phase with complete disorder characterized by an exponential decay of correlation at high noise amplitude. We have thoroughly investigated the interplay between the noise amplitude and the noise correlation time and thus obtained the complete, nonequilibrium stationary-state phase diagram in the relevant parameter space. Particular attention is provided on the dynamics of topological defects. We have observed that a finite correlation time promotes vortex excitations which is responsible for the decrease in the critical noise amplitude of the transition with an increase in correlation time.
As a special case, we recover the critical temperature of the equilibrium $BKT$ transition by studying a suitable limiting case of the dynamics~(\ref{eq:eom_DN}). The transition line in the phase diagram, when extrapolated, yields a well estimate of that equilibrium critical point.

We note that the introduction of dichotomous noise does not yield novel behavior in the sense that the nature of the ordered phase as well as order of the underlying transition remain same as observed in equilibrium dynamics of our system. Even when subject to Gaussian colored noise, the  nature of the transition remains unaltered. Despite that our analysis reveals some novel features associated with this nonequilibrium $BKT$-like transition, which are absent in white noise or colored noise driven system. These are the following.

Firstly, in presence of Gaussian colored noise, the dynamics in the thermodynamic limit  exhibits $BKT$ transition at critical temperature same as the equilibrium one \cite{paoluzzi2018effective}. In other words, introduction of finite correlation does not alter the $BKT$ transition temperature. But our analysis reveals that, in presence of dichotomous noise, the $BKT$-transition point indeed shifts on introduction of finite correlation.

Secondly, the dynamics in presence of Gaussian colored noise is expected to yield the maximum value of the power-law exponent of spatial correlation function to be same as that of equilibrium case i.e. $1/4$ \cite{paoluzzi2018effective}. We found in our study that this exponent exceeds the equilibrium upper bound due to nonequilibrium nature of the dynamics. Existence of such values ($>1/4$) of the exponent in the context of nonequilibrium $BKT$-like transition has already been reported in earlier works on 2D planar model \cite{luther1977critical,nelson1977universal} and recently in driven-dissipative condensates \cite{dagvadorj2015nonequilibrium,comaron2021non}. This implies that when subject to dichotomous noise, the quasi-ordered phase can sustain higher level of collective excitations leading to faster decay of the spatial correlation compared to Gaussian white or colored noise.

Apart from these two important differences, on a broader perspective, we note that our work is an example where Mermin-Wagner theorem\cite{mermin1966absence}, which essentially tells for an equilibrium system that a continuous symmetry can not be broken spontaneously at any finite temperatures in spatial dimensions two or lower, holds in a nonequilibrium system driven by dichotomous noise which is a non-Gaussian, discrete process. Finally, we want to emphasize on the fact that unlike equilibrium states that may all be characterized in terms of the well-founded Gibbs-Boltzmann ensemble theory encompassing microcanonical and canonical ensembles, until now there is no general tractable framework that allows to study nonequilibrium stationary states (NESS) on a common footing. This implies that NESSs need to be studied on a case-by-case basis. Thus, whether this underlying $BKT$-like transition is a general feature for all colored noise driven system is still an open question and requires an independent study. We believe our results put new insight in synchronization phenomena in locally coupled oscillator system subject to external field. In fact, the dynamics (\ref{eq:eom_DN}) can be thought of as zero temperature 2D $XY$ model in presence of an external field which is a dichotomous noise. So, an immediate interesting extension would be to obtain the resulting phase diagram in the context of 2D $XY$ model by adding Gaussian white noise in the dynamics. Investigations in this direction are going on and will be reported elsewhere.

\appendix
\section{Generation of Dichotomous noise}
\label{App_DN_noise_generation}
In our present study we consider the driving force $\zeta(t)$ to be a dichotomous random Markov process with equal transition rate $\lambda$ between the two states $\pm H$. We follow the method to generate the realizations of dichotomous noise as given in \cite{barik2006langevin}. We define the conditional probability $P(-H,t|x_0,t_0)$ (similarly, $P(H,t|x_0,t_0)$) to be the probability that the random force $\zeta(t)$ takes the value $-H$ (similarly, $H$) at time $t$, given that it was $x_0$ at some earlier time $t_0$. Here $x_0$ can have only two values $\pm H$. The corresponding master equations are given by
\begin{eqnarray}
	&&\frac{\rm d}{{\rm d}t} P(-H,t|x_0,t_0) = -\lambda P(-H,t|x_0,t_0) +  \lambda P(H,t|x_0,t_0), \nonumber \\
	\label{eq:master_eqn_DN}\\
	&&\frac{\rm d}{{\rm d}t} P(H,t|x_0,t_0) = \lambda P(-H,t|x_0,t_0) -  \lambda P(H,t|x_0,t_0).\nonumber
\end{eqnarray}

The conservation of total probability implies,
\begin{eqnarray}
	P(-H,t|x_0,t_0) +  P(H,t|x_0,t_0) = 1.
\end{eqnarray}
Solving the master equations (\ref{eq:master_eqn_DN}) with the initial condition $P(x,t|x_0,t_0) = \delta_{x,x_0}$ at $t=t_0$, we get,
\begin{eqnarray}
	&& P(-H,t|x_0,t_0) = \frac{1}{2} + \frac{1}{2} \left( \delta_{-H,x_0} - \delta_{H,x_0} \right) \exp\left[ -2 \lambda (t - t_0)\right], \nonumber \\
	\label{eq:master_eqn_soln_DN}\\
	&& P(H,t|x_0,t_0) = \frac{1}{2} - \frac{1}{2} \left( \delta_{-H,x_0} - \delta_{H,x_0} \right) \exp\left[ -2 \lambda (t - t_0)\right]. \nonumber
\end{eqnarray}
One can immediately check the consistency of the solutions by taking the limit $(t-t_0) \to \infty$ which corresponds to the stationary state. In this limit, one obtains, 
\begin{eqnarray}
	P_{\rm{st}}(-H) &&\equiv P(-H,\infty|x_0,t_0) \nonumber \\
	&&= \frac{1}{2} = P(H,\infty|x_0,t_0) \equiv  P_{\rm{st}}(H),
	\label{eq:master_eqn_soln_DN_st}
\end{eqnarray}
as expected.

Using the solutions (Eq.~\ref{eq:master_eqn_soln_DN}), we further obtain in the stationary state,
\begin{eqnarray}
	\langle \zeta(t) \rangle &&= 0, \nonumber \\
	\label{eq:DN_st_prop_from_mastereqn}\\
	\langle \zeta(t) \zeta(t')\rangle &&= H^2 \exp(-2\lambda|t-t'|).\nonumber	
\end{eqnarray}
One can thus identify the noise correlation time $\tau=1/2\lambda$. Thus using Eq.~\ref{eq:master_eqn_soln_DN} one can generate the realizations of dichotomous noise with amplitude $H$ and correlation time $\tau (= 1/2\lambda)$ in the following way:

Suppose, at time $t$ the random variable (force) is $-H$, at a later time instant $t_1=t+\Delta t (\Delta t \ll t)$ whether the force will switch to $H$ or remain at $-H$ is determined by the transition probability:
\begin{eqnarray}
	P(H,t_1|-H,t) = \frac{1}{2} - \frac{1}{2} \exp\left( -2 \lambda \Delta t \right).
	\label{eq:trans_prob1}
\end{eqnarray}

Now a uniformly distributed random number $a \in [0,1]$ is drawn and is compared against the above probability. If $a < P(H,t_1|-H,t)$, we accept the move and thus the force switches to the value $H$; otherwise we reject it and the force remains $-H$.

On the other hand, if at time $t$ the force is $H$, we calculate the transition probability to switch to $-H$ in the later instant $t_1=t+\Delta t$. This is given by
\begin{eqnarray}
	P(-H,t_1|H,t) = \frac{1}{2} - \frac{1}{2} \exp\left( -2 \lambda \Delta t\right).
	\label{eq:trans_prob2}
\end{eqnarray}
Again a uniformly distributed random number $a \in [0,1]$ is drawn and is compared against the above probability. If $a < P(-H,t_1|H,t)$, the force switches to the value $-H$; otherwise it remains at $H$. We note that the probabilities obtained in Eq.~(\ref{eq:trans_prob1}) and Eq.~(\ref{eq:trans_prob2}) are same. This is due to the fact that we have chosen equal transition rate between the two states.

Now, we update time to $t_1=t+\Delta t$ and repeat the above procedure for the next time instant $t_2 = t + 2 \Delta t$. We keep updating time and repeating the above procedure to generate sequence of dichotomous force $\zeta(t)$ switching between two values $\pm H$ with transition rate $\lambda = 1/2\tau$. We note that the time interval $\Delta t$ should be much smaller than the correlation time $\tau$.

We note that the dynamics in presence of dichotomous noise is piecewise deterministic. In numerical simulation, we generate the dichotomous force following the above way at each step of integration and integrate the governing dynamics of our system with fourth order Runge-Kutta method with integration time step $0.01$. The random number was generated using Mersenne Twister ($MT19937$) algorithm.

\section{To check stationarity}
\label{App_stationarity}
\begin{figure}[]
	\centering
	\includegraphics[scale=0.2]{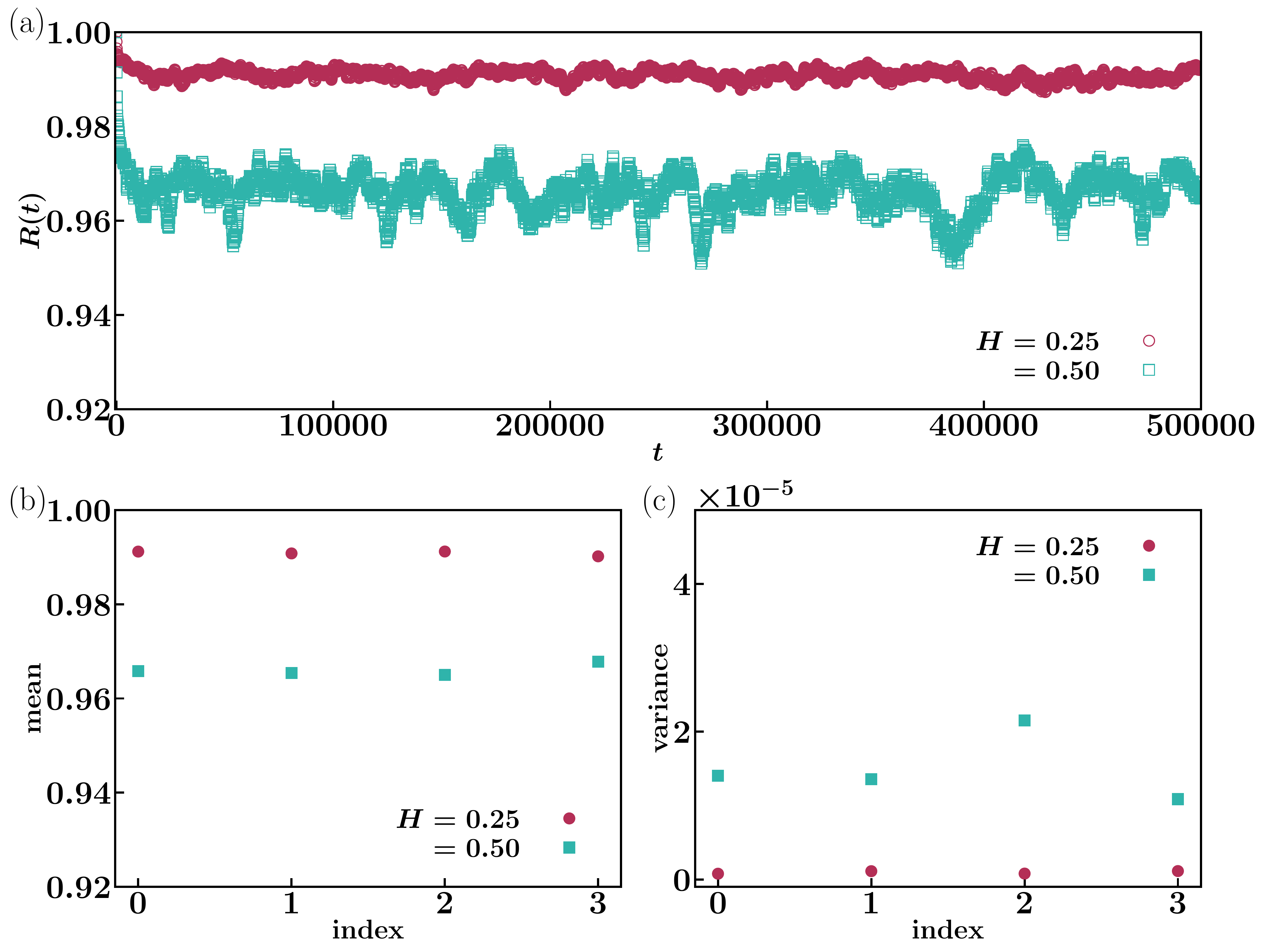}
	\caption{(a) Shown is time evolution of order parameter starting from an initial ordered state for two values of noise amplitude $H=0.25~\text{(empty circle)~and}~ 0.5$ (empty square), on a lattice of size $N=100 \times 100$. The noise correlation time is chosen to be $\tau=0.5$.  Here time $t$ is in integration time steps. Panels (b) and (c) display, for the same $H=0.25~\text{(filled circle)~and}~ 0.5$ (filled square), the mean and variance of the order parameter distribution over various nonoverlapping time windows (of length $t_w= 10^{5}$), respectively.}
	\label{fig:stationarity}
\end{figure}

To check whether the dynamics (\ref{eq:eom_DN}) reaches a stationary state, for a fixed $H$ and $\tau$, we initiate the dynamics from a synchronized phase on a fixed system size and record the value of order parameter $R$ (defined in Eq.~\ref{eq:order_para_definition}) with time. Figure \ref{fig:stationarity}(a) shows evolution of order parameter for two values of $H$ and fixed $\tau$, on a lattice of size $N=100 \times 100$. After an initial transient, we observe that $R$ does not change much in time. To check whether this is indeed stationary state or not, we divide the total time region discarding the initial transient ($t \geq 10^5$ time steps) in various nonoverlapping time windows of length $t_w=10^5$ and over each $t_w$, we compute the distribution of the order parameter $R$. To quantify it, we compute the mean and variance of the distribution for each $t_w$, as shown in  Fig.~\ref{fig:stationarity} (b) and Fig.~\ref{fig:stationarity} (c), respectively. We observe that they are approximately constant over all theses time windows, implying that the dynamics of the Kuramoto oscillators settled down to a stationary state. Now we further look at the distribution of noise over the same time windows, and obtain for each $t_w$ two discrete delta distributions of equal height situated at $\pm H$. This confirms that the noise also attains a stationary state\footnote{One may note that the dynamics of the noise is Markovian anyway. Thus if the system is initialized in such a way that $50 {\%}$ of oscillators are in the state $+H$ and $50 {\%}$ are in the state $-H$, it is not necessary to check for stationarity of the noise.}.

\section{Scaling behavior of the Binder cumulant}
\label{App_Binder_cumulant_scaling}
We first describe the scaling behavior of the Binder cumulant for a continuous phase transition in an equilibrium system and then will discuss its behavior at $BKT$ transition.
Equilibrium continuous phase transition is associated with a
singularity in the second derivative of the free energy of the system. It is a collective, cooperative behavior of a macroscopically large number of degrees of freedom and thus is observed strictly in an infinite system. Theoretically one can achieve the limit of an infinite system, but experiments and numerical analysis always deal with a system of finite size and in a finite system, the number of degrees of freedom is finite and thus everything is analytic.

Finite-size scaling (FSS) theory allows to estimate the critical point of phase transition, i.e., the parameter value at which a singularity occurs in an infinite system, by analyzing the data for large but finite systems. For our discussions of the finite-size scaling theory, consider a system exhibiting a continuous phase transition between two different phases characterized by a real scalar order parameter $\Psi$ as a function of temperature $T$: an ordered phase with $|\Psi|>0$ at temperatures below a critical temperature $T_c$ and a disordered phase characterized by $\Psi =0$ at and above $T_c$. We define $\epsilon \equiv (T-T_c)/T_c$.

Now we consider a system with linear dimension $L$ (so that $N$, the number of degrees of freedom, scales as $N \sim L^d$, with $d$ being the dimension of the embedding space) and denote the correlation length as $\xi(L)$ and the order parameter as $\Psi(L)$. Then, a continuous phase transition, observed as $L \to \infty$, is characterized by the divergence of the correlation length $\xi(\infty)$ at temperatures around the critical point as $\xi(\infty) \sim |\epsilon|^{-\nu};~\epsilon \to 0$, where $\nu$ is a critical exponent. The critical exponent $\beta$ \footnote{We want to convey to the readers that we use the notation $\beta$ for order parameter scaling exponent following standard convention. Remember that we used $\beta_{1}$ in a different context (as power law exponent of the temporal correlation) in the main text. These two are different.} characterizes the behavior of $\Psi(\infty)$ close to the critical point, as $\Psi(\infty) \sim (-\epsilon)^\beta;~\epsilon\to 0^-$. For large but finite $L$ and at a given $|\epsilon| \to 0$, if one has $L \gg \xi(\infty)$, no significant finite-size effects should be observed. On the other hand, for $L \ll \xi(\infty)$, the system size will cut-off long-distance correlations, and hence, one would expect finite-size rounding off of critical-point singularities. Thus one expects for small $\epsilon$ that the ratio $\xi(\infty)/L$ (or, equivalently, the ratio $|\epsilon|L^{1/\nu}$) controls the behavior of $\Psi$. So one may write under the assumptions of the finite-size scaling theory the following scaling form: 
\begin{eqnarray}
	&&\Psi(L) \sim L^{-\beta/\nu} f(|\epsilon|L^{1/\nu}).
	\label{eq:scaling-form} 
\end{eqnarray}
The scaling function $f(x)$, defined with $x>0$, satisfies the following properties:
\begin{eqnarray}
	f(x) &&\sim x^\beta, ~~~~~~ \text{as}~~ x\to \infty, \nonumber \\
	&&\to \text{constant.} ~~~~ \text{as} ~~x \to 0. \label{eq:fx_prop}
\end{eqnarray}
Such forms
ensure that as required, in the limit $L \to \infty$ at a fixed and small $\epsilon$, we
have $\Psi(\infty) \sim \epsilon^\beta$.
On the other hand, at a fixed $L$, as $T \to T_c$, one has $\Psi(L) \sim
L^{-\beta/\nu}$.

Binder cumulant $U(L)$\footnote{In our main text, we use the notation $U_{L}$ instead of $U(L)$ for convenience.} is defined from the estimates of the order parameter as follows \cite{binder1981finite},
\begin{eqnarray}
	U(L) \equiv 1- \frac{\langle (\Psi(L))^4 \rangle}{3\langle (\Psi(L))^2 \rangle^2}.
\end{eqnarray}	
For systems with continuous degrees of freedom in the limit $L \to \infty$, one has in the ordered phase the asymptotic behavior, $U(L) \to 2/3$, and in the disordered phase the asymptotic behavior, $U(L) \to 1/3$ \cite{binder1981finite,wysin2005extinction}. Now its behavior near the criticality can be understood as follows. 
One has similar to Eq.~(\ref{eq:scaling-form}) the scaling forms
\begin{eqnarray}
	&&\langle (\Psi(L))^2 \rangle \sim L^{-2\beta/\nu} f_1(|\epsilon|L^{1/\nu}), \nonumber\\
	&&\langle (\Psi(L))^4 \rangle \sim L^{-4\beta/\nu} f_2(|\epsilon|L^{1/\nu}).
	\label{eq:scaling-form2}	
\end{eqnarray}
with the scaling functions $f_1$ and $f_2$ having the same behavior
as the function $f$ in Eq.~(\ref{eq:fx_prop}). Consequently, we will have the scaling behavior 
\begin{eqnarray}
	&&U(L) \sim h(|\epsilon|L^{1/\nu}).
	\label{eq:scaling-form3}	
\end{eqnarray}
We thus observe that as $\epsilon \to 0$ at a fixed $L$, we have $U(L)=U^{*}$, a value that is $L$ independent. 

For large but finite $L$, one has in both the phases, the correlation length $\xi$ satisfying $\xi \ll L$, and consequently, $U(L)$ for various lattice sizes remains close to these aforementioned asymptotic values. Now, near criticality, the system is expected to stay close
to another fixed point value $U^{*}$ which is $L$-independent. Thus, the common intersection point of Binder cumulant curves for various system sizes yields an estimate of the critical point $T_c$.

\paragraph*{Near $BKT$ transition:}
Now for a system exhibiting a $BKT$ transition, in the region $T \leq T_c$, fluctuations diverge and consequently the correlation length $\xi(\infty)$ is infinite. Thus for large but finite $L$ the curves of $U(L)$ for various $L$ are expected to stay close to a fixed point $U^{*}$ in this region. In practical application, due to the statistical uncertainties, it is observed that $T_c$ is very close to the point where different curves begin to separate from the low-$T$ asymptotic value \cite{wysin2005extinction}.

For precise measurement of $T_c$ one needs to work with very large system-size and study along with the Binder cumulant other quantities e.g. the second moment correlation length and Helicity modulus simultaneously taking into consideration logarithmic corrections to them \cite{loison1999binder,hasenbusch2008binder}.

\paragraph*{FSS in Nonequilibrium systems:}
In a nonequilibrium system, in principle, one can not define free energy as these systems do not possess a Hamiltonian. But one can still define order parameter for such a system and thus compute quantities like Binder cumulant from the estimates of statistical averages of the order parameter. We further assume that the FSS forms for these quantities, defined in equilibrium systems, holds true in our nonequilibrium system too.

We note that Helicity modulus or spin-wave stiffness turns out to be a good candidate as order parameter for $BKT$ transition in equilibrium system. It quantifies the resistance of the system to a
twist in the boundary conditions, and is defined as the second derivative of the free-energy density of the system with a twist along one boundary axis \cite{fisher1973helicity}. One needs to know the jump in the Helicity modulus at criticality to estimate the transition point \cite{wysin2005extinction, loison1999binder, hasenbusch2008binder}. In our nonequilibrium system free energy can not be defined and we are unable to express the Helicity modulus in terms of statistical averages of the order parameter. In fact, there is no parameter analogous to temperature in our system. Having considered all these, we resort to the Binder cumulant only. Although the transition point estimated using Binder cumulant is imprecise, this approach reliably yields the nature of the ordered phase and thus the nature of the underlying transition. 

\section*{Data Availability}
The data that support the findings of this study are available from the corresponding author upon reasonable request.
\acknowledgments
I thank Neelima Gupte for introducing me to the field of synchronization in many-body interacting systems. I acknowledge very useful discussions on the $BKT$ transition with Mustansir Barma. I am grateful to Shamik Gupta for fruitful discussions and suggestions on the manuscript. I would also like to thank the anonymous referees for their valuable suggestions which helped us improve the manuscript. Finally, I thank HPCE, IIT Madras for providing me with high performance computing facilities in AQUA cluster. 

\bibliographystyle{aip}
\bibliography{References_mrinal}

\end{document}